\documentclass[hidelinks,12pt]{article}
\usepackage{amsmath}
\usepackage{graphicx,psfrag,epsf}
\usepackage{subcaption}
\usepackage{enumerate}
\usepackage[ruled,vlined]{algorithm2e}
\usepackage{natbib}
\usepackage{url} 
\usepackage{xcolor}
\usepackage{pdfpages}
\usepackage{setspace}
\usepackage{hyperref}
\usepackage{authblk}
\usepackage{soul}

\newcommand{\indep}{\rotatebox[origin=c]{90}{$\models$}}

\allowdisplaybreaks

\newtheorem{theorem}{Theorem}[section]

\newtheorem{corollary}[theorem]{Corollary}

\newtheorem{remark}{Remark}[section]
\newtheorem{definition}{Definition}[section]

\begin{document}
\date{}
\title{\bf Bayesian Robust Learning in Chain Graph Models for Integrative Pharmacogenomics}
\author[1]{Moumita Chakraborty}
    \author[2]{Veerabhadran Baladandayuthapani}
    \author[3]{Anindya Bhadra}
\author[1]{Min Jin Ha\thanks{
    Email: MJHa@mdanderson.org}}
\affil[1]{Department of Biostatistics, The University of Texas MD Anderson Cancer Center, Houston, TX}
\affil[2]{Department of Biostatistics, University of Michigan, Ann Arbor, MI}
\affil[3]{Department of Statistics, Purdue University, West Lafayette, IN}
\maketitle
\begin{abstract}
    Integrative analysis of multi-level pharmacogenomic data for modeling dependencies across various biological domains is crucial for developing genomic-testing based treatments. Chain graphs characterize conditional dependence structures of such multi-level data where variables are naturally partitioned into multiple ordered layers, consisting of both directed and undirected edges. Existing literature mostly focus on Gaussian chain graphs, which are ill-suited for non-normal distributions with heavy-tailed marginals, potentially leading to inaccurate inferences. We propose a Bayesian robust chain graph model (RCGM) based on random transformations of marginals using Gaussian scale mixtures to account for node-level non-normality in continuous multivariate data. This flexible modeling strategy facilitates identification of conditional sign dependencies among non-normal nodes while still being able to infer conditional dependencies among normal nodes. In simulations, we demonstrate that RCGM outperforms existing Gaussian chain graph inference methods in data generated from various non-normal mechanisms. We apply our method to genomic, transcriptomic and proteomic data to understand underlying biological processes holistically for drug response and resistance in lung cancer cell lines. Our analysis reveals inter- and intra- platform dependencies of key signaling pathways to monotherapies of icotinib, erlotinib and osimertinib among other drugs, along with shared patterns of molecular mechanisms behind drug actions.
\end{abstract}

\noindent%
{\it Keywords:} Bayesian graphical models; Cancer; Data integration; Robust graphical models; Multi-platform genomics; Pharmacogenomics.
\doublespacing
\newpage
\section{Introduction}\label{sec: intro}
 Pharmacogenomics encapsulates genomic mechanisms governing variable drug response and has been implemented into drug development pipeline to improve drug effectiveness and to reduce adverse drug reactions and toxicity \citep{squassina2010realities,Roden2019}. In cancer, utilizing the underlying genomic profile of tumors, especially the downstream effects of genes and their products for better understanding drug mechanisms, can lead to the development of more effective and robust treatment regimes \citep{kasarskis2011integrative}. In non-small cell lung cancer (NSCLC) patients, for example, clinical trials have shown that the epidermal growth factor receptor (EGFR) T790M mutation confers resistance to first-generation tyrosine kinase inhibitors (TKIs), which has led to the development of new EGFR inhibitors such as osimertinib that show efficacy superior to that of standard EGFR-TKIs \citep{mok2017osimertinib,soria2018osimertinib}. Patients with atypical EGFR mutations, on the other hand, show heterogeneous and reduced responses to EGFR inhibitors including osimertinib, and there are currently no established guidelines for the uncommon mutations \citep{robichaux2021structure}. The basic underlying premise is that accounting for the heterogeneity in drug sensitivity in relation to multiple molecular domains, while utilizing preclinical models of human cancer, is a key step toward discovering holistic functional mechanisms of anticancer drugs, which could facilitate better systems for classifying tumor and more robust clinical trial designs \citep{bedard2013tumour, lim2019emerging}.

These efforts have been catalyzed through consortium-level efforts such as the Cancer Dependency Map (The DepMap Portal; \url{www.depmap.org}), which provides a rich data repository for human cancer cell lines that encompass various types of primary cancers for identifying targetable genes and their functional relations across diverse domains of biological information. The portal includes multi-platform data such as copy number alteration (CNA), mRNA expression, and reverse phase protein array (RPPA) based protein expression obtained from the Cancer Cell Line Encyclopedia (CCLE) \citep{ghandi2019next}. It also contains drug sensitivity outcomes of more than $4,000$ drugs for these CCLE samples, which are based on high-throughput growth-inhibitory drug activities screened using Profiling Relative Inhibition Simultaneously in Mixtures (PRISM) technology \citep{corsello2020discovering}.
Most existing pharmacogenomic analyses are limited to identifying the association of molecular features with drug sensitivity without characterization of within- and cross-platform dependencies \citep{iorio2016landscape, corsello2020discovering}. A unified framework that provides a detailed characterization of multi-platform regulatory behavior can help in identifying key biological mechanisms of drug action to facilitate drug development.

\begin{figure}[t!]
    \centering
    \includegraphics[width=1.1\textwidth]{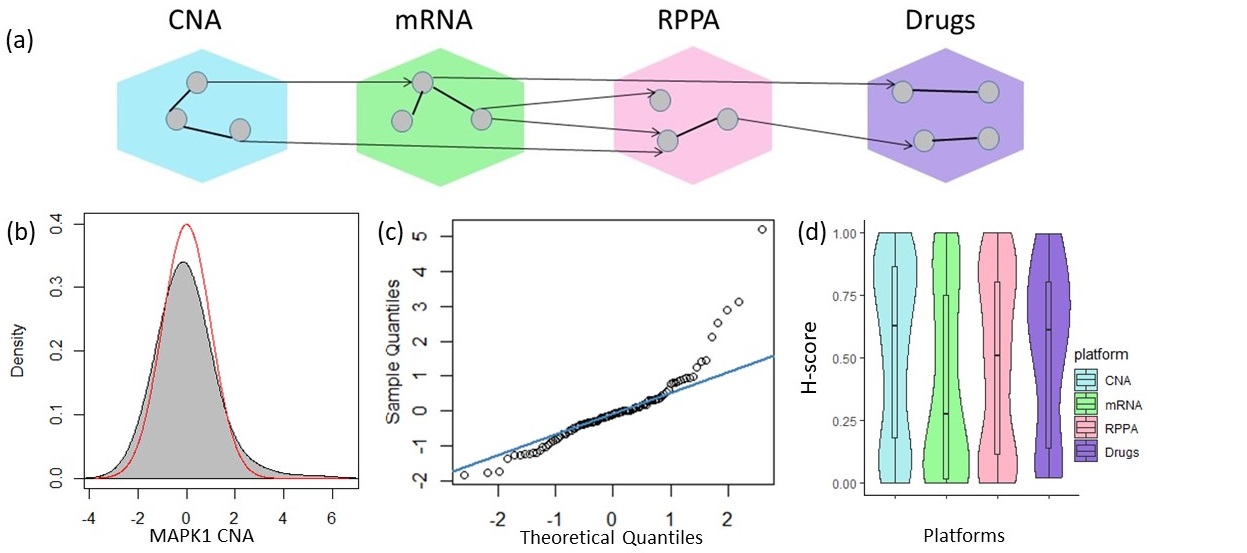}
    \vspace{-15pt}
    \caption{(a) Chain graph structure for CNA, mRNA, RPPA and drug layers, with directed and undirected edges between nodes of different and same layers respectively.(b) Empirical density plot of MAPK1 CNA levels. The $H$-score defined in the text as a measure of non-normality is equal to $0.988$ for MAPK1 CNA. The red curve is the density of standard normal distribution. (c) Normal q-q plot of data corresponding to MAPK1 CNA levels (d) $H$-scores across multi-platform genomic data and $20$ drugs. }
    \vspace{-10pt}
    \label{fig:plat_wise}
\end{figure}

To this end, we employ a graph-theoretic approach that captures dependencies among biological variables, where a node represents information on each unit and an edge between two nodes is an indicator of interaction or dependence between the corresponding biological units. For multi omic data, we assume that the entire set of nodes is partitioned into multiple platforms that are ordered based on fundamental biological principles. The platforms corresponding to datasets in our study are assumed to be ordered as: \{CNA $\rightarrow$ mRNA$\rightarrow$RPPA$\rightarrow$ Drugs\}, so that platforms lower in the hierarchy regulate data in higher platforms \citep{morris2017statistical}. This conceptual structure of our data is explained through Figure~\ref{fig:plat_wise}a, which can be cast as a \emph{chain graph} structure, where the set of nodes can be naturally partitioned into disjoint subsets, called layers with a pre-established order induced by biology. Edges between nodes within a layer are undirected and those between layers are directed, pointing towards the layer placed higher in the hierarchy. 

 Joint modeling of the mixed dependency structure in a chain graph, containing hierarchical multiple sub-graphs with directed and undirected edges, engenders substantial methodological and technical challenges. Most existing approaches rely on Gaussian assumptions on chain graphs, that simplify the methodological and computational complexities. Two-layer Gaussian chain graph models (GCMs) have been studied in the framework of multivariate Gaussian regressions or covariate adjusted Gaussian graphical models  \citep{rothman2010sparse,yin2011sparse,bhadra2013joint,chen2016asymptotically,li2021joint}. GCMs with more than two layers have been considered by \cite{LinMichailidis2016} through penalized maximum likelihood estimation of the coefficients and the precision matrix, and \cite{ha2020bayesian} who proposed a Bayesian approach to coherently learn chain graphs by variable selection on node-wise conditional likelihoods. More multi-layered GCM estimation methods can be found in \cite{drton2006maximum}, \cite{drton2008sinful}, \cite{mccarter2014sparse}, and \cite{petersen2018sparse} among others.

Although GCMs have been used (successfully) for a broad range of biological data that are continuous (or transformed to be continuous), they are ill-suited when the underlying variables exhibit considerable non-normal characteristics such as skewness, heavy tails and multimodal marginal distributions. 
As a motivating example, Figure~\ref{fig:plat_wise}b,c display the empirical density and normal quantile-quantile (Q-Q) plot respectively, of CNA levels of MAPK1, a gene controlling several cell signaling processes such as proliferation and transcriptional regulation in different cancers \citep{vicent2004mitogen}, across the $104$ lung cancer cell lines in our case study (detailed in Section~\ref{sec: application}). There is clear evidence of a heavier-than-normal tail due to extreme values observed in the right tail. We further quantify the ``degree of non-normality" for all nodes across the four layers, based on the score: $H(\boldsymbol x) = 2*\Phi(\log(1-pval(\boldsymbol x)))$,
where $\Phi$ is the cdf of standard normal distribution, and $pval(\boldsymbol x)$ is the \emph{p}-value of the Kolmogorov-Smirnov test for normality of $\boldsymbol x$. The $H$-score is between $0$ and $1$ with the higher value indicating the higher departure from normality. CNA and drug nodes show significant levels of non-normalities. Moreover, high levels of within-layer $H$-score variations are detected, which implies node-specific tail behavior. 
\vspace{-15pt}
\paragraph{Modeling background and the current state of the art.} Modeling non-normality in multivariate data has been performed using copulas \citep[e.g., ][]{nelsen2007introduction,genest2014modeling}. Gaussian copula models that use a set of latent variables following a multivariate normal distribution, have been discussed by \citep{pitt2006efficient,dobra2011copula}. \cite{liu2009nonparanormal} provide a semiparametric non-paranormal model and show that it is equivalent to a Gaussian copula when all the variables are continuous. Estimation methods for the non-paranormal model have been developed using various rank-based measures of dependence \citep{liu2012high, xue2012regularized}. 
These methods assume the transformation of the original variables into latent Gaussian variables to be deterministic. Relatively fewer works have focused on dependencies among variables under a random transformation.  

In networks with non-normal continuous marginals with random transformations leading to Gaussian latent variables, early Bayesian works include \cite{finegold2011robust} and \cite{finegold2014robust}, who modeled the node marginals using the multivariate-$t$ or Dirichlet-$t$ distributions. The multivariate-$t$ model of \cite{finegold2011robust} that assumes the same scaling transformation for each node can infer dependencies on the original variables through \emph{conditional uncorrelatedness}. This model is, however, less flexible than their alternative-$t$ model and the Dirichlet-$t$ that allow different node-wise scaling factors. To infer dependencies on the original scale of data using a more flexible model such as Dirichlet-$t$, \cite{bhadra2018inferring} proposed a Bayesian framework based on random scale transformations that helps in modeling skewed and heavy-tailed continuous marginals while allowing each node to have a different marginal distribution modeled in a data-dependent manner. Moreover, \cite{bhadra2018inferring} introduced the property of \emph{conditional sign independence}, which is weaker than conditional independence, but holds among \emph{observed variables} under non-normal marginals; we reserve formal definitions for Section~\ref{sec: robust model}. None of the above mentioned methods however, are applicable to chain graphs. We build on the approach of random scale-transformations in the context of non-Gaussian chain graphs, by allowing appropriately scaled transformations of node-marginals to be normally distributed, with the additional flexibility of inferring conditional (in)dependence on the observed data.  
\vspace{-35pt}
\paragraph{Summary of our novel contributions.} In this article, we develop a \emph{Robust Chain Graph Model (RCGM)} for multilayered non-Gaussian chain graphs that allows modeling of heavier-than-normal marginal tails in a flexible, data-dependent manner at each layer. Our approach makes multiple contributions:
\vspace{-10pt}
\begin{enumerate}
    \item \emph{Methodological contributions}: RCGM provides a highly practical way of bringing non-normality to large graphs by employing Gaussian scale transformations. Node-wise scaling factors are precisely calibrated using empirical marginal tail behaviors. Structural estimation of RCGM uses selection priors on the edges that induce sparsity in the network. RCGM therefore offers the computational advantages of fast high-dimensional graph estimation methods, in addition to accurate adjustments for heavy-tailed marginals in chain graphs.
    \vspace{-10pt}
    \item \emph{Theoretical contributions}: RCGM introduces a new Markov property of conditional sign independence that allows for node-specific non-normalities in a chain graph and interprets dependencies on the original scale of variables (Theorem~\ref{th: condl sign indep}). Edges between normal nodes can still be interpreted in terms of conditional independence.
    \vspace{-35pt}
    \item \emph{Scientific contributions}: RCGM addresses the growing need in cancer biology research for delineating  multiplatform functional drivers of networks underlying drug action. RCGM is designed to handle data needed for such analyses, which come from multiple platforms and often deviate significantly from normality. In our study, RCGM is used to infer the dependency structure within and between CNA, mRNA, RPPA and drug data platforms in lung cancer cell lines. Our analysis encompasses key signaling pathways in human cancers, revealing pathway-level genomic features regulating drug responses. Notable findings include the overall receptivity of the DNA Damage Response (DDR) pathway to drugs, particularly associations of the protein levels of Checkpoint Kinase 1 (CHK1), the main effector of DDR, with drug actions of EGFR TKIs erlotinib and icotinib, which have considerable clinical utility. 
\end{enumerate}
\vspace{-10pt}
The rest of the paper is organized as follows. In Section~\ref{sec: robust model} we introduce the robust chain graph model. Section~\ref{sec: estimation} describes the Bayesian structural estimation procedure. Section~\ref{sec: simulations} provides a comparison of the performance of RCGM with existing GCM alternatives. In Section~\ref{sec: application}, we analyze the data involving four biological platforms of lung cancer cell lines, to investigate driving mechanisms of drug sensitivities on major lung cancer drugs. Proofs and additional relevant details are presented in the Supplementary Materials.  The R codes for the RCGM implementation and data are also available in the supplemental files.
\vspace{-15pt}
\section{Model}\label{sec: robust model}
\vspace{-10pt}
\noindent \emph{Data structure and notations.} We consider independent and identically distributed data across cell lines, with $\boldsymbol X=(X_1, \ldots , X_q)^T$ denoting data corresponding to a cell line comprising of $q$ coordinates. The $q$ nodes can be partitioned into $L$ disjoint subsets, each subset to be called a layer. Due to existing biological hierarchies, the layers possess an inherent ordering among themselves and are numbered following that order, such that data in higher layers are regulated by data in lower layers. For instance, in our case study, $L=4$ and the ordering from layer 1 to 4 is: CNA$<$mRNA$<$RPPA$<$drugs. The dependency structure is modeled by a chain graph, with directed and undirected edges between nodes belonging to different and same layers, respectively. Let $G=(V, E, \mathcal{L})$ denote the chain graph over $q$ nodes, where $V=\{1, \ldots , q\}$ is the set of labeled nodes across $L$ layers, $E$ is the set of directed ($\rightarrow$) and undirected (--) edges between nodes in $V$, and $\mathcal{L}$ is a mapping of node indices to their corresponding layers so that for $v\in V$, $\mathcal{L}(v)$ is the index of the layer in which $v$ is located. For layer $l$, let $\boldsymbol X_{(l)}$ be the sub-vector of $\boldsymbol X$ with nodes in layer $l$. Let $\boldsymbol X_{[1:l]}$ denote the sub-vector of nodes in layers $1$ to $l$. Let $\mathcal{T}_l$ be the set of nodes in layer $l$ and $q_l=\sum_{j=1}^l|\mathcal{T}_l|$, where $|A|$ is the number of elements in the set $A$. Depending on the probability distribution of $\boldsymbol X$, edges can be characterized in different ways.
\vspace{-15pt}
\subsection{Gaussian Chain Graph Models}
\vspace{-10pt}
A Gaussian chain graph model (GCM) is a multilayered graph where $\boldsymbol X$ follows a multivariate Gaussian distribution. It uses an undirected Gaussian graph to model within layer interactions and a block recursive normal linear simultaneous equations-model to describe dependencies of higher layers on lower layers. This way, the full model is composed of layer-wise normal regression components described by:
\vspace{-20pt}
\begin{align}\label{def: mlGGm formulation 2}
    \boldsymbol X_{(1)}  \sim N_{q_1}(\boldsymbol 0, \mathcal{J}_1^{-1}), \quad\text{and}\quad \boldsymbol X_{(l)} | \boldsymbol X_{[1:l-1]} \sim N_{|\mathcal{T}_l|}(\boldsymbol \beta_l\boldsymbol X_{[1:l-1]}, \mathcal{J}_l^{-1}), \quad l\geq 2,
\end{align} \par
\vspace{-15pt}
where $\boldsymbol \beta_l$ and $\mathcal{J}_l$ are respectively coefficient and precision matrices for the $l$-th regression component. The model therefore consists of $L-1$ multiple regressions and one undirected graph in the first layer. Nonzero entries of $\boldsymbol \beta_l$ and $\mathcal{J}_l$ respectively encode directed and undirected edges in GCM $G$. Specifically, $(u-v)\in E$ when the entry corresponding to $(u,v)$ in $\mathcal{J}_l$ equals zero for nodes $u$ and $v$ in the same layer $l$. Similarly, $(u\rightarrow v)\in E$ when $\boldsymbol\beta_l$'s entry for $(v, u)$ is zero, for $\mathcal{L}(u)<\mathcal{L}(v)$ and $\mathcal{L}(v)=l$. By Remark 4.1 and Theorem 4.1 of \cite{andersson2001alternative}, the chain graph $G$ given by \eqref{def: mlGGm formulation 2} follows the alternate Markov property (AMP) where edges are characterized by conditional independence given the AMP conditioning nodes (detailed in Section~\ref{sec: Characterization of dependencies in RCGM}), when the true distribution of $\boldsymbol X$ is multivariate normal. Estimation methods for GCM can be found in \cite{drton2006maximum}, \cite{mccarter2014sparse}, \cite{LinMichailidis2016}, \cite{petersen2018sparse} and \cite{ha2020bayesian}, among others. 

Although GCMs capture dependencies in multi-level continuous multivariate data, they are inappropriate in settings where marginal distributions are heavy-tailed so that the joint multivariate distribution is no longer Gaussian. An example of a heavy-tailed undirected graph to model such data is the multivariate \emph{t}-distribution discussed by \cite{finegold2011robust}. In chain graphs with heavy-tailed marginals, using a Gaussian model can result in incorrect inferences \citep{genest2014modeling}, and dependencies given by zero structures of $\boldsymbol \beta_l$ and $\mathcal{J}_l$ cannot be interpreted in terms of conditional independence based on the AMP in  \cite{andersson2001alternative}. By modeling the marginal non-normal behaviors of nodes, we resolve these issues in a new framework of robust chain graph models that can accurately infer the network and yield interpretable notions of dependencies.
\vspace{-15pt}
\subsection{Robust Chain Graph Models (RCGM)}
\vspace{-10pt}
An effective way to incorporate non-normality arising due to heavy tails is by scale mixture representations with appropriate factors so that the transformed data follow a multivariate Gaussian distribution. Let $d_v$ denote the positive scaling factor for node $v$ such that the $d_v$, $v\in V$ are independent and have $d_v\sim p_v$ for positive scaling distributions $p_v$ with $\int d_vp_v(d_v)<\infty$. Let $\boldsymbol D$ be a $q\times q$ diagonal matrix with entries $(1/d_1, 1/d_2, \ldots , 1/d_q)$, $d_v>0$. Given $d_v$'s, the transformed data $\boldsymbol D\boldsymbol X$ is assumed to follow a multivariate normal distribution. These random scale transformations have been used in the context of single-layer undirected graphs \citep{finegold2011robust,finegold2014robust,bhadra2018inferring}. In our framework of robust chain graphs, we allow flexibility in the marginal behavior through scale transformations with node-specific degrees of tail-heaviness.

Let $\boldsymbol D_l$ and $\boldsymbol D_{[1:l]}$ denote the sub-matrices of $\boldsymbol D$ corresponding to the nodes $(q_{l-1}+1)$ to $q_l$ and from $1$ to $q_l$ respectively. We define our Robust Chain Graph Model (RCGM) as
\vspace{-20pt}
\begin{align}\label{eq: nn-chain graph building}
    \boldsymbol D_l\boldsymbol X_{(l)} &= \boldsymbol B_l\boldsymbol D_{[1:l-1]}\boldsymbol X_{[1:l-1]}+ \boldsymbol \varepsilon_l,\quad \varepsilon_l\sim \mathrm{N}_{|\mathcal{T}_l|}(\boldsymbol 0, \mathcal{K}_l^{-1}),\quad 2\leq l\leq L,\nonumber\\
     \varepsilon_1 &=\boldsymbol D_1\boldsymbol X_{(1)}, \quad \varepsilon_1 \sim\mathrm{N}_{q_1}(\boldsymbol 0, \mathcal{K}_1^{-1}),
\end{align} \par
\vspace{-20pt}
where $\mathcal{K}_l$ is the $|\mathcal{T}_l|\times |\mathcal{T}_l|$ precision matrix for the transformed data in the $l$-th layer and $\boldsymbol B_l$ is a $|\mathcal{T}_l|\times q_{l-1}$ coefficient matrix, and the layer-specific error vectors $\boldsymbol\varepsilon_l$ are independent of each other. The independence of $\varepsilon_l$'s ensures that the conditional distribution of $\boldsymbol D\boldsymbol X$ given $\boldsymbol D$ is a multivariate normal distribution given by $\boldsymbol D\boldsymbol X \big|\boldsymbol D \sim \mathrm{N}_q(\boldsymbol{BDX}, \mathcal{K}^{-1})$, 
where $\mathcal{K}$ is a $q \times q$ precision matrix with entries $k_{uv}$, $\boldsymbol B$ is a $q \times q$ coefficient matrix, and the sub-matrix of $\boldsymbol B$ corresponding to rows $q_{l-1}+1$ to $q_l$ and columns $1$ to $q_{l-1}$ is equal to $\boldsymbol B_l$, while that of $\mathcal{K}$ for rows $q_{l-1}+1$ to $q_l$ and columns $q_{l-1}+1$ to $q_l$ is $\mathcal{K}_l$. So given $\boldsymbol D$, we have $\boldsymbol{DX}=\boldsymbol{BDX}+\boldsymbol\varepsilon$, $\boldsymbol\varepsilon\sim\mathrm{N}_q(\boldsymbol 0, \mathcal{K}^{-1})$, so that $\boldsymbol{DX}\big | \boldsymbol D  \sim \mathrm{N}_q\left(\boldsymbol 0, \Omega^{-1}\right)$, 
    $\Omega = (\boldsymbol I - \boldsymbol B)^T\mathcal{K}(\boldsymbol I - \boldsymbol B)$.

Note that the RCGM in (\ref{eq: nn-chain graph building}) is GCM on the transformed set of variables $\boldsymbol D\boldsymbol X$, and includes GCM as a special case for unit scaling factors for all nodes. When the true distribution is non-normal, dependencies in the $E$ obtained from GCM or RCGM are no longer determined in terms of AMP conditional independence of \cite{andersson2001alternative}. In Section~\ref{sec: Characterization of dependencies in RCGM}, we characterize dependencies in the RCGM using a weaker Markov property.
\vspace{-15pt}
\subsection{Characterization of Dependencies in RCGM}\label{sec: Characterization of dependencies in RCGM}
\vspace{-10pt}
The edge set $E$ may hold different statistical interpretations of dependencies, depending on the true probability distribution of the chain graph. The RCGM as described in \eqref{eq: nn-chain graph building} is constructed in a way to satisfy AMP of \cite{andersson2001alternative} in a Gaussian population. The AMP specifies a direct mode of data generation and provides an easily interpretable statistical characterization of directed and undirected edges in chain graph models. In non-normal populations, however, RCGM does not satisfy the AMP, as nonzero entries of $(\boldsymbol B,\mathcal{K})$ may not imply conditional independence. Nevertheless, edges in RCGM can be characterized in terms of the dependencies defined as follows:
\vspace{-5pt}
\begin{definition}
We define four types of relations between random variables $Y_1$ and $Y_2$ with a conditioning random vector $\boldsymbol Z$.\\
(i) $Y_1$ and $Y_2$ are said to be \ul{conditionally sign-independent (CSI)} given $\boldsymbol Z$, denoted by $Y_1\indep^s~Y_2|\boldsymbol Z$, if $\mathrm{P}(Y_1 < 0| \boldsymbol Z) = \mathrm{P}(Y_1 <0 |Y_2, \boldsymbol Z)$ and $\mathrm{P}(Y_2 < 0| \boldsymbol Z) = \mathrm{P}(Y_2 <0 |Y_1, \boldsymbol Z)$, whenever the conditional probabilities exist. Otherwise, $Y_1$ and $Y_2$ are \ul{conditionally sign-dependent (CSD)} given $\boldsymbol Z$.\\
(ii) $Y_1$ and $Y_2$ are said to be \ul{conditionally independent (CI)} given $\boldsymbol Z$, denoted by $Y_1\indep Y_2|\boldsymbol Z$, if $f(Y_1| \boldsymbol Z) = f(Y_1|Y_2, \boldsymbol Z)$ and $f(Y_2| \boldsymbol Z) = f(Y_2 |Y_1, \boldsymbol Z)$, where $f$ denotes the corresponding probability density functions, whenever the conditional probability densities exist. Otherwise, $Y_1$ and $Y_2$ are \ul{conditionally dependent (CD)} given $\boldsymbol Z$.
\end{definition}
\vspace{-5pt}
As a simple illustration, consider $Y_1\sim N(0,1)$, $Y_2=UY_1$, $Z\sim N(0,1)$, where $U\sim\text{Unif}(-1,1)$, and $U$ and $Z$ are independent of $Y_1$. Then $(Y_1,Y_2)$ are conditionally sign-independent given $Z$. The concept of CSI in Definition 2.1 (i) is defined in terms of the sign rather than the magnitude, which was introduced in \cite{bhadra2018inferring} for undirected networks, implying that sign of $Y_1$ is independent from that of $Y_2$, conditioned on $\boldsymbol{Z}$. Thus, CSI is weaker than CI. However, the advantage is that it applies to less restrictive multivariate models where the normality assumption can be relaxed.

\begin{figure}[t!]
    \centering
    \includegraphics[width=0.7\textwidth]{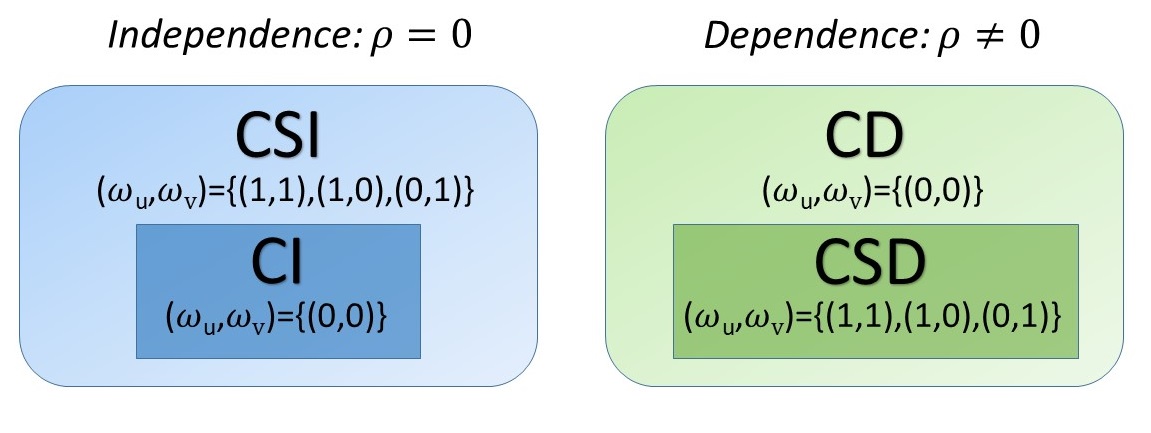}
    \vspace{-20pt}
    \caption{The nested relation between CSI, CI, CD and CSD (Definition 2.1). For $\mathcal{L}(u)=\mathcal{L}(v)$, $\rho=\mathcal{K}_{uv}$. For $\mathcal{L}(u)<\mathcal{L}(v)$, $\rho=\boldsymbol B_{vu}$. $\omega_v$ is the indicator of $v$ being non-normal. For missing edges ($\rho=0$) in $G$, CI is the stronger form of independence as compared to CSI while for edges ($\rho\neq 0$) in $G$, CSD is stronger than CD in terms of dependence.}
    \vspace{-10pt}
    \label{fig: sign indep}
\end{figure}

In our RCGM model~\eqref{eq: nn-chain graph building} that allows both normal and non-normal nodes, we aim to characterize the edges in terms of CSI and CI  on the original random variables $\boldsymbol{X}$. We model the node-specific non-normality behavior by introducing indicators $\omega_v$ that takes the value $1$ if $X_v$ has tails heavier than normal ($d_v\sim p_v$) and $0$ if $X_v$ is normal ($d_v=1$).
Edge interpretations can be obtained from Theorem~\ref{th: condl sign indep}.
\vspace{-10pt}
\begin{theorem}\label{th: condl sign indep}
(i) (At least one node is non-normal). If $\omega_u=1$ or $\omega_v=1$, conditional sign-independence follows from $\boldsymbol B$ and $\mathcal K$ as:
\vspace{-10pt}
\begin{enumerate}
    \item[(a)] ($u$ and $v$ in the same layer). Suppose $\mathcal{L}(u)=\mathcal{L}(v)$ and $\rho=k_{uv}=k_{vu}$. Then $\rho=0$ if and only if $X_u\indep^s~X_v|\boldsymbol Z_u$, where $\boldsymbol Z_u =\boldsymbol X_{[1:\mathcal{L}(u)]}\backslash\{X_u, X_v\}$.
    \vspace{-10pt}
\item[(b)] ($u$ and $v$ in different layers). Suppose $\mathcal{L}(u)<\mathcal{L}(v)$ and $\rho=\boldsymbol B_{vu}$. Then $\rho=0$ if and only if $X_u\indep^s~X_v|\boldsymbol Z_d$, where $\boldsymbol Z_d=\boldsymbol X_{[1:\mathcal{L}(v)-1]}\backslash X_u$.
\end{enumerate}
\vspace{-10pt}
(ii) (Between normal nodes). Suppose $\omega_u=\omega_v=0$, and $\rho$ is as defined in part (i). Then $\rho = 0$ if and only if $X_u\indep X_v|\boldsymbol Z_u$ for $\mathcal{L}(u)=\mathcal{L}(v)$ and $X_u\indep X_v|\boldsymbol Z_d$ for $\mathcal{L}(u)<\mathcal{L}(v)$.
\end{theorem}
\vspace{-5pt}
The proof is in Supplementary Section~\ref{sec: supp. proof 2.1. part 1} - \ref{sec: supp. proof 2.1. part 2}. Note that the conclusion (ii) of Theorem~\ref{th: condl sign indep} is the same as the AMP of \cite{andersson2001alternative}. Theorem~\ref{th: condl sign indep} shows that the interpretation of $E$ given by RCGM depends on the node-specific marginal distributions of the corresponding random variables. The types of dependencies and the way they are related to each other are demonstrated in Figure~\ref{fig: sign indep} and summarized as follows: 
\vspace{-10pt}
\begin{itemize}
    \item \emph{CSI and CI.} A missing edge between two nodes is interpreted as conditional sign-independence (CSI) when at least one of the nodes is non-normal by part (i) of Theorem~\ref{th: condl sign indep}. When both nodes follow normal distributions, by part (ii) of the theorem, the absence of an edge between two nodes is interpreted as conditional independence (CI) which is a stronger form of the Markov property.
    \vspace{-10pt}
    \item \emph{CSD and CD.} The non-zero entries of $\boldsymbol B$ and $\mathcal{K}$ are interpreted as conditional sign-dependence (CSD) which implies conditional dependence (CD) when at least one of the corresponding nodes is non-normal. Thus, edges connecting non-normal nodes are considered to have stronger relation than edges between normal nodes.
    \vspace{-10pt}
\end{itemize}

Another notion of dependency, the partial correlation, is defined as the correlation between variables after adjusting for the effects of conditioning variables, and the zero values are equivalent to conditional independence in a Gaussian population. However, as the zero structure is preserved under random and independent marginal scale transformations (see Section 5.1 of \cite{finegold2011robust}), partial correlation interpretations of edges do not hold in non-Gaussian populations, as explained in Remark~\ref{remark: partial correlation}. 
\vspace{-10pt}
\begin{remark}\label{remark: partial correlation}
Neither of the edge sets $E$ obtained under GCM or RCGM can be interpreted in terms of partial correlations between original variables under non-normal data, as evaluating partial correlations involves the true probability distribution. RCGM yields CSI interpretations of $E$, which is the main advantage of using RCGM over GCM in the non-normal network estimation scenario.
\end{remark}
\vspace{-25pt}
\subsection{Incorporating a Continuous Measure of Non-normality}
\vspace{-10pt}
To allow for more flexibility in the proposed framework, we incorporate a mixture model for the amount of non-normality by introducing a variable $\pi_v$ for node $v$ which quantifies the departure of the node's distribution from that of a normal distribution. We treat the non-normality status $\omega_v$ of each node as random, and model $P(\omega_v=1) = \pi_v$, a priori, so that the larger $\pi_v$ indicates a higher degree of prior belief regarding departure from normality for node $v$.  We formulate our final model as
\vspace{-15pt}
\begin{align}\label{def: mlGGm formulation final}
    \boldsymbol D\boldsymbol X| \boldsymbol D, \boldsymbol\omega, \boldsymbol\pi \sim N(\boldsymbol{BDX}, \mathcal{K}^{-1}), \quad
    d_v|\boldsymbol\omega, \boldsymbol\pi \sim \omega_v p_v + (1-\omega_v)\delta_1,\quad
    \omega_v|\pi_v &\sim \text{Bern}(\pi_v),
\end{align} \par
\vspace{-15pt}
where $\pi_v\in [0, 1]$ for all $v\in V$ and $\boldsymbol\pi= (\pi_1,\ldots, \pi_q)$. 
By considering $\omega_v$ as random under model \eqref{def: mlGGm formulation final}, the strength of dependencies are interpreted in a continuous scale calibrated by probabilities rather than deterministically as in Theorem 2.1. 

Model selection-based network inference methods commonly assign a probability score $g_{uv}$ to each edge, which indicates its probability of being present in the network. The edge-wise probability score $g_{uv}$ and node-wise non-normality scores $(\pi_u,\pi_v)$ are combined to characterize the dependence properties for a network, by the following corollary to Theorem~\ref{th: condl sign indep}, with a proof in Supplementary Section~\ref{sec: supp. proof 2.2}.
\begin{corollary}\label{th: chain graph condl sign indep 3}Let $g_{vu}$ be the probability of edge $(u - v)$ or $(u\rightarrow v)$ being present in the model under $\mathcal{L}(u)=\mathcal{L}(v)$ and $\mathcal{L}(u)<\mathcal{L}(v)$ respectively. Then $X_u\indep^s~X_v|\boldsymbol Z$ with probability $(1-g_{vu})$ and $X_u\indep X_v|\boldsymbol Z_u$ with probability at least $(1-g_{vu})(1-\pi_u-\pi_v+\pi_u\pi_v)$ for $\boldsymbol Z=\boldsymbol X_{[1:\mathcal{L}(u)]}\backslash\{X_u, X_v\}$ when $\mathcal{L}(u)=\mathcal{L}(v)$ and $\boldsymbol Z=\boldsymbol X_{[1:(\mathcal{L}(v)-1)]}\backslash X_u$ when $\mathcal{L}(u)<\mathcal{L}(v)$.
\end{corollary}

Corollary~\ref{th: chain graph condl sign indep 3} provides a probabilistic characterization of edges in RCGM on the original scale of nodes, from CSI to CI for missing edges, and equivalently CD to CSD for edges, depending on the probabilities of edge inclusion and node non-normality. In summary, the multilayered network inference using RCGM in \eqref{def: mlGGm formulation final} allows a robust structural recovery of chain graphs when the marginals deviate from normal distributions, along with calibration of node-specific marginal tail-heaviness through $\pi_v$, and dependence characterization of the network structures on the original variables weighted by the strength from CSD to CD. 
\vspace{-15pt}
\section{Bayesian Estimation of RCGM}\label{sec: estimation}
\vspace{-10pt}
Estimation of the multilayered network incorporating uncertainty in node-wise normality involves two parameters $\boldsymbol B$ and $\mathcal{K}$ for the graphical structure and $\boldsymbol\pi$ for the degree of non-normality in model~\eqref{def: mlGGm formulation final}. We use a Bayesian framework with a Markov chain Monte Carlo (MCMC) sampling scheme to draw posterior samples of the parameters. Since the RCGM is formulated in layer-wise multivariate regressions (Equation~\eqref{eq: nn-chain graph building}), we perform each regression by imposing priors independently across layers. At each MCMC iteration, for each layer $l$, we update the structural parameters $\boldsymbol B_l$ and $\mathcal{K}_l$, following a scale transformation of the data using sampled scaling factors with $\boldsymbol \pi_l$. For Sections~\ref{sec: Prior Calibration of Node-wise Non-normality} -- \ref{sec: posterior sampling}, let $\boldsymbol x_v$ and $\boldsymbol d_v$ respectively denote the $n\times 1$ vectors of data $\boldsymbol X$ and scaling matrix $\boldsymbol D$ corresponding to node $v$.
\vspace{-15pt}
\subsection{Prior Calibration for Node-wise Non-normality}\label{sec: Prior Calibration of Node-wise Non-normality}
\vspace{-10pt}
The prior specifications of node non-normality are conducted based on empirical marginal distributions. We assume that $\pi_v$ follows a beta distribution
\vspace{-20pt}
\begin{align*}
    \pi_v &\sim \text{Beta}(\mu_vr_v, (1-\mu_v)r_v),
\end{align*} \par
\vspace{-20pt}where $\mu_v$ is the prior mean and $r_v$ is to control the variance. The mean and variance are decided based on the degree of non-normality evaluated from data for each node. Specifically, we set $\mu_v$ by the $H$ score defined as $H(\boldsymbol x_v) = 2*\Phi(\log(1-pval(\boldsymbol x_v)))$ from the \emph{p}-value of the test for normality of the marginal distribution of $X_v$ from the Kolmogorov-Smirnov test. We choose $r_v$ to ensure a small variance (e.g., 0.01) for the prior distribution to be concentrated around $H(\boldsymbol x_v)$ on the unit interval.

The mixing distribution $p_v$ in the model~\eqref{def: mlGGm formulation final} is determined by empirically evaluating the tail behavior of marginals for each node. The marginal tail mass appearing as exponential or polynomial decay is related to exponential or polynomial tail behavior of $p_v$ respectively \citep{bhadra2018inferring}. We assume that every heavy-tailed marginal centered at its median is a univariate-\emph{t} or a double exponential distribution with non-centrality or a location parameter of zero respectively, so that there are two categories of $p_v$ - polynomially decaying such as an Inverse-Gamma, and exponentially decaying as in the Gamma and Exponential distributions. We regress $\log \hat{f}(\boldsymbol x_v)$, the log-transformed smoothed empirical probability density estimate of the marginal distribution, on $\log \boldsymbol x_v$ and $\boldsymbol x_v$. We then determine for which category (polynomial/exponential) the regression \emph{p}-value is smaller and estimate the coefficients for $\log \boldsymbol x_v$ and $\boldsymbol x_v$ in this category. The estimated coefficients are then used to derive the parameters of the chosen $p_v$ using Algorithm~\ref{alg: alg 1} in Supplementary Section~\ref{sec: selection of mixing dist}.
\vspace{-15pt}
\subsection{Priors on Model Selection Parameters}\label{sec: Priors on Model Selection Parameters}
\vspace{-10pt}
 The scaling factors $d_v$ are generated using the non-normality parameters $\pi_v$ from Section 3.1. As RCGM becomes a GCM for the scaled variables when $d_v$ are given (Equation~\eqref{eq: nn-chain graph building}), $\boldsymbol B$ and $\mathcal{K}$ can be estimated based on the scaled data by building Gaussian and Wishart priors for the layer-wise regression parameters $\boldsymbol B_l$ and $\mathcal{K}_l$. However, in chain graphs with multiple large layers, the set of parameters becomes extensive with the number and size of layers, in addition to becoming increasingly sparse \citep{armstrong2005bayesian}. Significant computational challenges appear even in moderately large graphs. 
Instead, we simultaneously select undirected and directed edges connected to a node $v$ belonging to layer $l$, using a stochastic search variable selection framework (SSVS) \citep{george1993variable} after coherently reparameterizing the precision parameter $\mathcal{K}_l$ into regression coefficients to yield the node-conditional likelihoods \citep{ha2020bayesian}. For $v$, the node-wise regression is
\vspace{-15pt}
\begin{align}\label{eq: node-wise transform 2}
        \frac{X_v}{d_v} &=(\boldsymbol X\boldsymbol D)^T_{[1:l-1]}(\boldsymbol b_v  - \boldsymbol B_l^{(v)}\boldsymbol a_v)+ (\boldsymbol X\boldsymbol D)^T_{\mathcal{T}_l\backslash v}\boldsymbol a_v + e_v,
    \end{align} \par
\vspace{-15pt}
where $\boldsymbol a_v=-k_{vv}\boldsymbol k_l^{(v)}$, where $\boldsymbol k_l^{(v)}$ is the vector of $k_{vu}$, $u\in \mathcal{T}_l\backslash v$, and $e_v\sim N(0, k_{vv}^{-1})$. The parameters of interest in each node-conditional likelihood are $\boldsymbol b_v$, $\boldsymbol a_v$, $\boldsymbol B_l^{(v)}$, and $k_{vv}$, where the effect of nodes in layers $1:(l-1)$ on $v$ is depicted by $\boldsymbol b_v$, while their effect on nodes in $\mathcal{T}_l\backslash v$ is denoted by $\boldsymbol B_l^{(v)}$, and $\boldsymbol a_v$ is the effect of the neighbors of node $v$ in $\mathcal{T}_l$ on $v$. Details can be found in Supplementary Section~\ref{sec: prior formulation supp.}. However, as $\boldsymbol b_v$, $\boldsymbol a_v$, and $\boldsymbol B_l^{(v)}$ are not jointly identifiable (Supplementary Section~\ref{sec: identifiability}), we fix $\boldsymbol B_l^{(v)}$ at its current value in each MCMC iteration and consider the node-conditional likelihood with parameters $\boldsymbol b_v$, $\boldsymbol a_v$ and $k_{vv}$. 

We set priors on the parameters of each node-wise regression for estimating $\boldsymbol B_l$ and $\mathcal{K}_l$ in the model~\eqref{def: mlGGm formulation final}. As shown in the equation (10) of \cite{ha2020bayesian}, Wishart and independent Gaussian priors on $\boldsymbol B_l$ and $\mathcal{K}_l$ respectively are equivalent to independent Gaussian and Gamma priors on $\{\boldsymbol b_v,\boldsymbol a_v\}$ and $k_{vv}$. We let $\gamma_{vw}$ and $\eta_{vu}$ be the indicator variables encoding zero-structures of $\boldsymbol B$ and $\mathcal{K}$, with $P(\gamma_{vu}=1) =p_{vw}$, and $P(\eta_{vu}=1)=q_{vu}$ for $p_{vw},q_{vu} \in (0, 1)$. We use a spike-and-slab prior similar to \cite{ha2020bayesian} to set priors as
\vspace{-20pt}
\begin{align*}
    b_{vw}|\gamma_{vw},k_{vv} &\sim \gamma_{vw}N(0, c^2_{vw}/ k_{vv})+(1-\gamma_{vw})\delta_0, \\
    a_{vu}|\eta_{vu}, k_{vv} &\sim \eta_{vu}N(0, 1/(\lambda_lk_{vv})) + (1-\eta_{vu})\delta_0,\\
    k_{vv} &\sim Gamma((\delta_l+|\mathcal{T}_l|-1)/2, \lambda_l/2),
\end{align*}\par
\vspace{-20pt}
for $c_{vw}, \lambda_l, \delta_l>0$ and $\delta_0$ denotes the degenerate distribution at $0$. The MCMC algorithm is run layer-wise at each iteration, wherein the scaled node-wise likelihoods corresponding to nodes in $\mathcal{T}_l$ (in random order of nodes) are combined with the priors to yield posterior samples for $\boldsymbol b_v$, $\boldsymbol a_v$ and $k_{vv}$ keeping $\boldsymbol B_l^{(v)}$ fixed.  

 \begin{algorithm}[t]
\SetAlgoLined
 At iteration $t$, update $\boldsymbol D$ given current $\boldsymbol \pi$ for every subject by Metropolis-Hastings sampling using Equation~\eqref{eq: MCMC generate di}.\\
 \For{$1\leq l\leq q_L$}{
  \For{$v \in \mathcal{T}_l$}{
  \textup{Update $\pi_v$ using Metropolis Hastings sampling (Equation~\eqref{eq: MCMC update pi})\\}
    \textup{Update the undirected edges:}\textup{~ 
    \begin{enumerate}
    \item Set response $\tilde{\boldsymbol x}_v=\boldsymbol x_v/\boldsymbol d_v - \boldsymbol X_{[1:l-1]}^T\boldsymbol D_{[1:l-1]}\boldsymbol b_v$ and covariates $\boldsymbol Z_v$ as $\boldsymbol D_{\mathcal{T}_l\backslash v}\boldsymbol X_{\mathcal{T}_l\backslash v} - \boldsymbol B_{\mathcal{T}_l\backslash v,[1:l-1]}\boldsymbol D_{[1:l-1]}\boldsymbol X_{[1:l-1]}=\boldsymbol\epsilon_{\mathcal{T}_l\backslash v}$.
        \item Update $\{\boldsymbol\eta_w:w\in\mathcal{T}_l\}$ by an add/delete/swap step and MH-selection thereafter. \\Set the neighborhood of $v$ as $\text{ne}_v^{u}=\{w\in\mathcal{T}_l:\eta_{vw}\neq 0\}$. Set $\eta_{wv}=\eta_{vw}$ for every $w\in \text{ne}_v^{u}$.
        \item Update the coordinates of $\boldsymbol  a_v$ that were selected into $\text{ne}_v^{u}$ and $ k_{kk}$ using Gibbs sampling through Equations~\eqref{eq: sampling gibbs alpha 1} and \eqref{eq: sampling gibbs kappa 1}.
    \end{enumerate}}
    \textup{Update the directed edges:}\textup{~
    \begin{enumerate}
        \item Set $\tilde{\boldsymbol x}_v= \boldsymbol x_v/\boldsymbol d_v - \boldsymbol \epsilon_{\mathcal{T}_l\backslash v}^T\boldsymbol a_v$, $\boldsymbol Z_v= \boldsymbol D_{[1:l-1]}\boldsymbol X_{[1:l-1]}$, where $\boldsymbol \epsilon_{\mathcal{T}_l\backslash v}$ is equal to $\boldsymbol D_{\mathcal{T}_l\backslash v}\boldsymbol X_{\mathcal{T}_l\backslash v} - \boldsymbol B_{\mathcal{T}_l\backslash v,[1:l-1]}\boldsymbol D_{[1:l-1]}\boldsymbol X_{[1:l-1]}$.
       \item Update $\{\boldsymbol\gamma_w:w\in\mathcal{P}_l\}$ using add/delete/swap and MH-selection. Set $\text{ne}_v^{d}=\{w\in\mathcal{P}_l:\gamma_{vw}\neq 0\}$, where $\mathcal{P}_l$ is the set of nodes in layers $1$ to $l-1$.
        \item Update the coordinates of $\boldsymbol b_v$ that were selected into $\text{ne}_v^{d}$ and $ k_{vv}$ using Gibbs sampling through Equations \eqref{eq: sampling gibbs alpha 2} and \eqref{eq: sampling gibbs kappa 2}.
    \end{enumerate}}
  }
 }
 \caption{\label{alg: alg mcmc} MCMC sampling steps for iteration $t$.
}
\end{algorithm}
\vspace{-15pt}
 \subsection{Posterior Sampling} \label{sec: posterior sampling}
 \vspace{-10pt}
 Posterior samples of $\boldsymbol B$, $\mathcal{K}$ and $\boldsymbol\pi$ are constructed from the parameters $\boldsymbol b_v$, $\boldsymbol a_v$, $k_{vv}$ and $\pi_v$ in node-wise regressions (Equation~\eqref{eq: reparametrize back B K supp}), which are drawn using the MCMC sampling scheme summarized in Algorithm~\ref{alg: alg mcmc}, with detailed derivations in Supplementary Section~\ref{supp:mcmc}. A maximum a posteriori (MAP) estimate across the MCMC samples may be hard to derive in huge model spaces, and provides no probabilistic quantification of uncertainty in the parameters. Instead, we use the marginal posterior edge inclusion probability $g_{uv}$, which is the proportion of times in MCMC runs after burn-in that the edge ($u-v$ or $u\rightarrow v$) is included. Fixing the false discovery rate (FDR) $\alpha$ in $(0, 1)$, we determine a cutoff $C_\alpha$ by sorting all $g_{uv}$ in decreasing order to obtain $g_{(t)}$, and setting $C_\alpha=g_{(\xi)}$, where $\xi=\max\{k:k^{-1}\sum_{t=1}^k(1-g_{(t)})<\alpha\}$. We then form the set of edge discoveries $\chi_\alpha=\{(u,v):g_{uv}>C_\alpha\}$. We similarly evaluate the sign of an edge by the sign of coordinate-wise average of $\boldsymbol B$ or $\mathcal{K}$, and the non-normality probabilities $\pi_v$, using the corresponding averages across MCMC samples. We combine posteriors for the node non-normalities ($\pi_u$, $\pi_v$) and the edge inclusion probability $g_{uv}$ to assign weight for an edge between nodes $u$ and $v$ that represents the strength of dependence from CSD to CD based on Corollary~\ref{th: chain graph condl sign indep 3}.
 \vspace{-15pt}
\section{Simulations}\label{sec: simulations}
\vspace{-10pt}
We conduct simulation experiments to evaluate the performance of our RCGM framework, in terms of graph structure recovery, as compared to other GCM-based methods, under various non-normality mechanisms. We generate simulation datasets based on the model in Equation~\eqref{def: mlGGm formulation final} corresponding to random chain graphs with $q$ nodes that are divided into $L$ ordered layers with similar sizes. The layer-wise undirected graphs are formed by randomly connecting two nodes with probability $p_E$ independent of all other edges. We then connect two nodes in different layers independently with probability $p_E/2$, where the directions follow the order among the $L$ layers. Thus directed edges between layers are less likely to be connected than the undirected edges within a layer. Given the chain graph, we set the corresponding nonzero elements of $\boldsymbol B$ and $\mathcal{K}$ by random samples from a uniform distribution in $(-1.5, -0.5)\bigcup(0.5, 1.5)$ and ensure positive definiteness of $\mathcal{K}$ by imposing diagonal dominance. 
We consider pre-fixed non-normality score $\pi\in[0,1]$ and scale-mixing distribution $p_v$ for coordinate $v$ for each simulation dataset. For every $\pi$ and $p_v$ combination, we impose node-wise heavy tails on each sample by generating  $\omega_v\sim\text{Bernoulli}(\pi)$, $d_v\sim p_v$ if $\omega_v=1$, $d_v=1$ otherwise and then transforming the sample as $(X_1,\ldots , X_q) \mapsto (X_1d_1, \ldots , X_qd_q)$. We repeat this process for all the $n$ samples generated from the Gaussian chain graph model. We consider mixing distributions with two types of tails - exponential with $p_v$ as the exponential distribution with mean $2.5$, and polynomial where $p_v$ is Inverse-Gamma with shape $3$ and scale $6$. We vary $\pi$ across a range of values in $(0,1)$ corresponding to low ($\pi=0.05$), medium ($\pi=0.60$) and high ($\pi=0.95$) levels of non-normality, expecting RCGM to  perform better than GCMs for datasets with higher $\pi$. 

We compare RCGM's performance with that of the Bayesian node-wise Gaussian approach \citep{ha2020bayesian} and the penalized Gaussian likelihood approach \citep{LinMichailidis2016}, using BANS and LBBM respectively to refer to these methods. We use $4,000$ burn-in samples and $10,000$ samples for posterior inference in both RCGM and BANS, and determine the cutoffs on edge inclusion posterior probabilities in both methods by controlling the FDR at $0.1$. As suggested by \cite{LinMichailidis2016}, we use glasso \citep{friedman2008sparse} to estimate the undirected graph in the first layer by LBBM. Table~\ref{tab: simulation results} displays the performance of the three methods across three values of $\pi$ in terms of the metrics described as $Specificity= TN/(TN+FP)$, $Sensitivity= TP/ (TP+FN)$, and Matthew's correlation coefficient $MCC=[(TP\times TN) - (FP\times FN)]/[\{ (TP+FP)(TP+FN)(TN+FP)(TN+FN) \}^{1/2}]$ that ranges from -1 (complete non-concordance) to 1 (full concordance). Across various tuning parameters for LBBM and cutoffs for the posterior probability of edge inclusion for BANS and RCGM, the performance is evaluated based on area under the ROC curve (AUC). We calculate the partial area under ROC curve (pAUC) by evaluating the area under the curve between specificity ranging from $0.8$ or $0.9$ to $1$ and dividing it by the maximum possible $AUC$ in that range.

\begin{table}[h!]
    \centering
    \rule{460pt}{1pt}\\
    $(q,L, p_E)=(50,4, 0.08)$,  $p_v: \text{Exponential}(\text{mean}=2.5)$
    \resizebox{\columnwidth}{!}{\begin{tabular}{cc|cccccc}
    Setting & Method & Specificity & Sensitivity & MCC  & AUC & pAUC 0.9 & pAUC 0.8\\
    \hline
       & RCGM  &  \textbf{0.965} (0.006)
 & \textbf{0.812} (0.068) & \textbf{0.759} (0.046) & \textbf{0.902} (0.036) & \textbf{0.755} (0.013) & \textbf{0.795} (0.016)\\
        ($\pi=0.95$) & BANS & 0.922 (0.005) & 0.768 (0.088) & 0.701 (0.062) & 0.868 (0.042) & 0.684 (0.011) & 0.788 (0.016)\\
         & LBBM  & 0.905 (0.005) & 0.672 (0.088) & 0.655 (0.064) & 0.874 (0.041) & 0.731 (0.013) & 0.760 (0.016)\\
        \hline
         & RCGM  & \textbf{0.948} (0.007)
& \textbf{0.838} (0.064) & \textbf{0.787} (0.052)& \textbf{0.916} (0.038)
& \textbf{0.752} (0.035) & \textbf{0.791} (0.091)\\
        ($\pi=0.60$) & BANS & 0.939 (0.007) & 0.783 (0.077) & 0.738 (0.063) & 0.851 (0.044) & 0.735 (0.017) & 0.780 (0.022)\\
       & LBBM  & 0.940 (0.005) & 0.780 (0.088) & 0.731 (0.064) & 0.858 (0.039) & 0.738 (0.013) & 0.780 (0.016) \\
       \hline
        & RCGM  & 0.945 (0.007)
& \textbf{0.850} (0.061) & \textbf{0.821} (0.055)& \textbf{0.962} (0.021)
& \textbf{0.828} (0.017) & \textbf{0.874} (0.022)\\
        ($\pi=0.05$) & BANS & \textbf{0.955} (0.009) & 0.848 (0.072) & 0.820 (0.063) & 0.945 (0.027) & 0.820 (0.035) & 0.871 (0.091)\\
       & LBBM  & 0.912 (0.005) & 0.802 (0.088) & 0.792 (0.061) & 0.919 (0.042) & 0.795 (0.013) & 0.851 (0.016)
    \end{tabular}}
    \rule{460pt}{1pt}\\
 $(q,L, p_E)=(50, 4, 0.08)$,  $p_v: \text{Inverse-Gamma}(\text{shape}=3, \text{scale}=6)$
 \resizebox{\columnwidth}{!}{\begin{tabular}{cc|cccccc}
    Setting & Method & Specificity & Sensitivity & MCC  & AUC & pAUC 0.9 & pAUC 0.8\\
    \hline
     & RCGM  & 0.995 (0.007)
& 0.657 (0.064) & \textbf{0.604} (0.052)& \textbf{0.873} (0.044)
& \textbf{0.706} (0.035) & \textbf{0.748} (0.091)\\
       ($\pi=0.95$) & BANS & \textbf{0.996} (0.007) & 0.528 (0.077) & 0.553 (0.063) & 0.833 (0.046) & 0.698 (0.017) & 0.734 (0.022)\\
       & LBBM  & 0.882 (0.019) & \textbf{0.657} (0.041) & 0.557 (0.059) & 0.845 (0.052) & 0.683 (0.043) & 0.745 (0.036)\\
        \hline
         & RCGM  & \textbf{0.996} (0.006)
& 0.642 (0.078) & \textbf{0.610} (0.056)& \textbf{0.902} (0.029)
& \textbf{0.770} (0.032) & \textbf{0.781} (0.089)\\
        ($\pi=0.60$) & BANS & 0.996 (0.009) & 0.557 (0.075) & 0.527 (0.068) & 0.812 (0.052) & 0.726 (0.018) & 0.754 (0.025)\\
       & LBBM  & 0.879 (0.036) & \textbf{0.657} (0.052) & 0.535 (0.071) & 0.831 (0.051) & 0.723 (0.074) & 0.770 (0.076) \\
       \hline
        & RCGM  &  0.990 (0.006)
 & 0.757 (0.081) & \textbf{0.604} (0.051) & \textbf{0.961} (0.025) & \textbf{0.805} (0.020) & \textbf{0.840} (0.018)\\
        ($\pi=0.05$) & BANS & \textbf{0.994} (0.007) & 0.748 (0.091) & 0.598 (0.062) & 0.952 (0.027) & 0.803 (0.017) & 0.832 (0.019)\\
        & LBBM  & 0.877 (0.032) & \textbf{0.771} (0.044) & 0.583 (0.036) & 0.901 (0.034) & 0.773 (0.059) & 0.804 (0.061)\\
        \hline
    \end{tabular}}
    \vspace{-5pt}
    \caption{Performance of RCGM as compared to BANS and LBBM under Exponential and Inverse-Gamma scaling distributions $p_v$, and low, medium and high levels of non-normality indexed by $\pi=0.05, 0.60, 0.95$ respectively, and $(q, L, n, p_E)=(50, 4, 200, 0.08)$ where $q$, $L$, $n$ and $p_E$ denote the dimension of the graph, number of layers, sample size and sparsity respectively. pAUC 0.9 is the (scaled) area under the ROC curve when specificity is fixed at $0.9$ (1-specificity is fixed at $0.1$). Results are summarized across 30 replications; standard errors are within parentheses. The winning entry for each metric is displayed in bold.}
    \vspace{-10pt}
    \label{tab: simulation results}
\end{table}

 \begin{figure}[!h]
 \includegraphics[width=\textwidth]{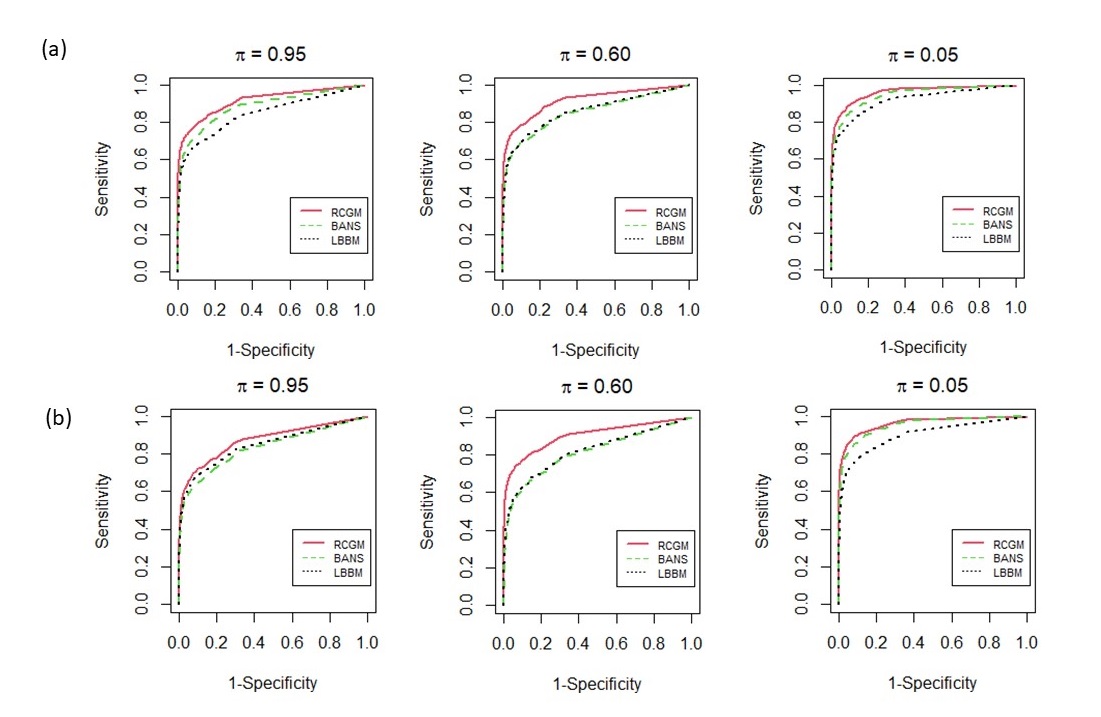}
 \vspace{-25pt}
     \caption{ROC curves for the simulation setting $(q, L, n, p_E)=(50, 4, 200, 0.08)$ across high, medium and low levels of non-normality $\pi$, where $q$, $L$ and $p_E$ denote the dimension of graph, number of layers and sparsity respectively. Panels (a) and (b) correspond to scaling by $\text{Exponential}(\text{mean}=2.5)$ and $\text{Inv-Gamma}(\text{shape}=3,\text{rate}=6)$ respectively.}
     \vspace{-10pt}
     \label{fig:edgepth11-4}
 \end{figure}
 \vspace{-10pt}
 We observe that performance of all the methods deteriorates gradually as more non-normality is induced through increasing $\pi$. While the three methods almost coincide for near-normal data when $\pi=0.05$, RCGM performs significantly better than the Gaussian methods for data with medium and high degree of non-normality (Table 1). 
 Therefore RCGM performs as good as GCMs in normal data, and consistently better than GCMs as tails become heavier than normal. Analysis of ROC curves reveals a relatively weaker performance of all the methods for medium non-normality ($\pi=0.60$) as compared to high non-normality ($\pi=0.95$), and a bigger contrast between RCGM and GCM methods in medium $\pi$ as compared to that in high $\pi$ (Figure~\ref{fig:edgepth11-4}).
 Simulations over an extended set of $\pi$ values with $\pi=\{0.01, 0.05, 0.4, 0.6, 0.8, 0.9, 0.95, 0.99\}$ show a similar pattern (Figure~\ref{fig:AUCvsPi}), where the AUCs are lower for $\pi=0.4$ to $\pi=0.8$ than for $\pi\geq 0.9$, and the maximum contrast between RCGM and GCM AUCs is found in the medium non-normality range $\pi\in[0.4,0.8]$. The possible reason behind this could be the higher level of tail-heaviness in the data induced by $\pi$ in the range $\pi=0.2$ to $\pi=0.8$. Our algorithm is designed to tackle data with heavy-tailed marginals, so the difference between performances of RCGM and Gaussian methods becomes more prominent with increased tail-heaviness, which occurs in the range $\pi=0.2$ to $\pi=0.8$. Further details can be found in Supplementary Section~\ref{sec: supplementary simulation details}.
\vspace{-15pt}
\section{Pharmacogenomics in Lung Cancer}\label{sec: application}
\vspace{-10pt}
Integrative data analysis and the use of network topology towards functional characterization of drug sensitivity is critical to the successful development of cancer treatments \citep{kasarskis2011integrative}. Our aim is to understand mechanisms of drug action by modeling the complex regulatory and interactive processes across various domains of the molecular data. A public resource for high-throughput screening data on more than 4,000 drugs for $578$ cell lines spanning $24$ human tumor types was created by \cite{corsello2020discovering} from the CCLE project \citep{barretina2012cancer}. Growth inhibitory activity defined as drug sensitivity was measured in terms of log-transformed median fluoroscence intensity (MFI) of barcoded cell lines after drugs were administered. Lower log-MFI values correspond to lower cell viability and therefore higher drug sensitivity. We use these drug screening data along with copy number aberration (CNA), mRNA expression (mRNA), and RPPA-based protein expression (RPPA) \citep{ghandi2019next} obtained from the DepMap Portal (\url{www.depmap.org}).
We select $n=104$ lung cancer cell lines and match them across the CNA, mRNA, RPPA and drug platforms. Sixteen drugs that have been evaluated in clinical trials for different types of lung cancer, are found in the PRISM database and are selected for the study. Based on the mechanism of action, these drugs can be categorized into EGFR-TKIs, ALK-TKIs, tubulin polymerization inhibitors among several other categories (Tables~\ref{tab: drug mech1} -- \ref{tab: drug mech2}). Four of these drugs - Cisplatin, Sevoflurane, Carboplatin and Sorafenib - that have been studied for potential combination therapies for NSCLC \citep{liang2013effects,langer1995paclitaxel, gridelli2011sorafenib}, are also selected to explore their functional mechanisms. Features in each platform constitute a layer, and the order CNA$<$mRNA$<$RPPA$<$drugs is justified by the biological process that CNA affects mRNA gene expressions, which are then translated into protein, and genes and proteins regulate drug response as a phenotype \citep{morris2017statistical}. 

We perform pathway-wise analyses that define multi-platform functional cancer networks for each pathway, based on literature  outlining abnormalities of cell signaling pathways as etiology of cancers including lung \citep{vogelstein2004cancer,brambilla2009pathogenesis}. We select genes/proteins that are involved in the 10 most clinically targetable signaling pathways in human cancers, as defined in \cite{akbani2014pan}. The gene and antibody of RPPA membership for each pathway is provided in Table~\ref{tab:antibody-gene pathway wise list}. We apply RCGM on each pathway-level multilayered data, using a burn-in sample size of $2,000$ and $10,000$ samples for posterior inference of edge-inclusion probabilities $g_{vu}$ and non-normality scores $\pi_v$. Controlling FDR at $0.1$, we selected edges that have posterior probabilities of edge inclusion greater than 0.55 across all pathways. 

The size of the parameter space in each pathway-level network is fairly large - the average number of nodes and number of parameters for edge inclusion and non-normality probabilities has an average of $46.2$ and $1090.32$ across pathways respectively. The computation time is reasonable, as the average time taken for estimating each network is $4.32$ hours with a standard deviation of $0.86$ across pathways, on a 3.5
GHz Intel Core i7 processor.

In our robust multilayered networks, the edges are weighted and colored by the dependence characterization of the RCGM discussed in Sections 2.3-2.4. Two nodes $u$ and $v$ are connected and weighted by probability $g_{vu}$, to represent the strength from CD to CSD based on Corollary~\ref{th: chain graph condl sign indep 3}. With node labels for non-normal marginals if $\hat\pi_v>0.5$, we further categorize the edges as CD if both are normal and CSD otherwise. An inter-platform connectivity analysis across pathways is displayed in Figure~\ref{fig: sankey diagram}, and the pathway-level multilayered networks are displayed in Figure~\ref{fig: DDR graph} and Figures~\ref{fig:graph_apoptosis} -- \ref{fig:graph_corereactive}. We further evaluate the extent of non-normality in the data and find high variability in within-platform non-normal behavior (Figure~\ref{fig:Estimated pi violin plot}), similar to the $H$-scores for empirical non-normalities (Figure~\ref{fig:plat_wise}d). The posterior non-normality probabilities $\hat{\pi}_v$ are positively correlated with H-scores (Figure~\ref{fig:nnscore_cellcycle}). 
\vspace{-40pt}
\subsection{Inter-platform Regulatory Network}
\vspace{-5pt}
\begin{figure}[t!]
    \centering
    \includegraphics[width=\textwidth]{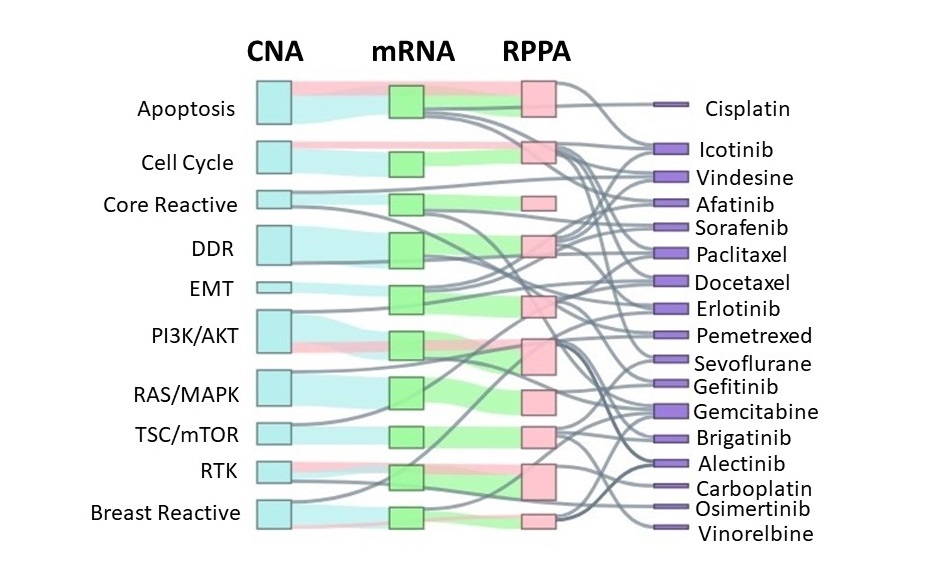}
    \vspace{-35pt}
    \caption{Sankey diagram showing connectivity between the $4$ platforms across $10$ pathways. Each box in the left three columns is a pathway-molecular platform combination, and widths of the lines between them are proportional to the number of directed edges connecting them. Gray lines denote edges between pathway-platform blocks and drugs.}
    \vspace{-10pt}
    \label{fig: sankey diagram}
\end{figure}
Based on the estimated networks across all 10 pathways, we investigate  inter-platform regulatory signaling patterns at the pathway-level in the Sankey diagram (Figure~\ref{fig: sankey diagram}). It demonstrates the connectivity between platforms within each pathway, and the drugs. Each unit is a pathway-platform combination depicted by a box and the lines between units are proportional to the number of directed edges between platforms within pathways. Sizes of the unit boxes are proportional to the degree, and larger boxes, therefore, represent higher levels of regulatory signaling coming in and out the pathways and drugs. 
The total number of directed edges between molecular platforms are $61$ for CNA$\rightarrow$mRNA, $56$ for mRNA$\rightarrow$RPPA and $13$ for CNA$\rightarrow$RPPA, indicating that as expected, the immediate platforms are tightly connected across pathways. Apoptosis and PI3K/AKT pathways have the most cross-platform signaling, with 8, 6, and 4 edges for CNA$\rightarrow$mRNA, mRNA$\rightarrow$RPPA and CNA$\rightarrow$RPPA respectively in Apoptosis, and 8, 7, and 3 edges respectively in PI3K/AKT. We investigate the regulatory factors to the drugs- CNA, mRNA and RPPA have $8$, $10$, and $17$ edges connected to any of the drugs, which implies that proteins are the most relevant factors that directly affect drug sensitivity. This is expected as protein kinases serve as crucial targets for drug development \citep{davies2006point}. Proteins in the cell cycle pathway show the most connectivity with drugs across pathway-wise RPPA at $4$ directed edges (Figure~\ref{fig:graph_cellcycle}). RPPA levels of Cyclin B1, E1 and E2 are found to regulate drug actions; these are cyclin-dependent kinase (CDK) inhibitors in the cell cycle that hold key significance in lung cancer cell proliferation \citep{baldi2011tumor}. Gemcitabine has the maximum number of connections with genomic platforms across drugs, with $3$ edges from mRNA and RPPA of CAV1 in core and breast reactive pathways (Figure~\ref{fig:graph_breastreactive} -- \ref{fig:graph_corereactive}). Dependence of gemcitabine on CAV1 and its products has been explored in cell line studies and clinical trials which have shown that CAV1 over-expression can lead to gemcitabine-resistance in lung cancer cells \citep{ho2008caveolin,shi2020multifaceted}. 
\vspace{-15pt}
\subsection{Multilayered Pathway-level Networks} 
\vspace{-10pt}
\begin{figure}[t!]
    \centering
    \includegraphics[scale=0.5]{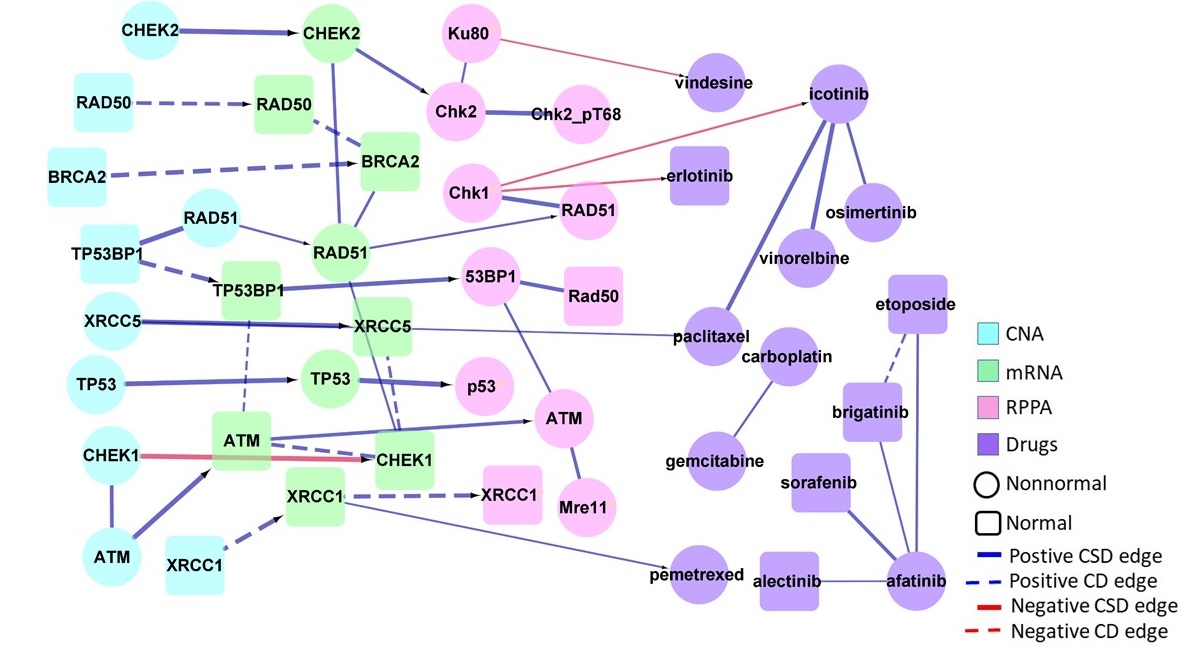}
    \vspace{-10pt}
    \caption{The estimated multilayered network for DNA Damage Response pathway. Blue and red edges indicate positive and negative dependencies, while CD and CSD stand for conditionally dependent and conditionally sign-dependent edges respectively. The width of the edges is proportional to the posterior inclusion probabilities.}
    \vspace{-10pt}
    \label{fig: DDR graph}
\end{figure}
Inter-platform connectivity analysis from Figure~\ref{fig: sankey diagram} shows that the DNA damage response (DDR) pathway has the highest level of cross-platform signaling with drugs. Dysregulation of DDR promotes mutations that lead to chemotherapy resistance in lung cancer, and the understanding and characterization of inter- and intra- platform molecular mechanisms that affect drug sensitivity are crucial to the development of targeted therapies in lung cancer \citep{burgess2020therapeutic}.
DDR molecular platforms are connected to drugs through $5$ edges, with $1$, $1$, and $3$ edges to drugs from CNA, mRNA and RPPA respectively (Figure~\ref{fig: DDR graph}). There are $10$ and $6$ edges respectively from CNA$\rightarrow$mRNA and mRNA$\rightarrow$RPPA, while no edges from CNA$\rightarrow$RPPA. We observe negative dependencies of cell viability on protein levels of Checkpoint Kinase 1 (CHK1), the main effector of DDR, when cells are administered with EGFR TKIs erlotinib and icotinib, indicating high sensitivity of CHK1 toward these two drugs. Further, the CHK1 protein has a positive dependency with RAD51 protein expression which has cis-acting regulatory elements at the mRNA and CNA levels. Sensitivity towards pemetrexed and paclitaxel are found to be dependent on mRNA expressions of DNA repair genes XRCC1 and XRCC5, whose polymorphisms may affect DNA repair capacity and thus regulate cancer progression \citep{schneider2008xrcc1}. 

Another interesting observation is on the positive dependencies between EGFR CNA, mRNA and protein levels and the sensitivity of EGFR CNA to osimertinib (Figure~\ref{fig:graph_rtk}). Higher levels of EGFR CNA are found to be associated with faster cell death when administered with osimertinib. EGFR sensitivity to osimertinib can be explained by clinical trials that show osimertinib successfully targets EGFR-mutant variants of NSCLC and shows improved efficacy over mutation-resistant standard EGFR-TKIs and platimun-based chemotherapies \citep{mok2017osimertinib, soria2018osimertinib}. 

Several pairs of drug-drug dependencies appear in most of the pathways-level networks (Figure~\ref{fig: drug-drug interactions 0}). For instance, the positive dependency between EGFR-TKIs icotinib and osimertinib is present in all the pathways. Both these drugs target EGFR, and are therefore expected to increase or decrease cell life span through similar patterns.
 \vspace{-20pt}
\section{Discussion}\label{sec: discussion}
\vspace{-10pt}
In this article, we develop a multilayered network estimation framework, Robust Chain Graph Model (RCGM), to estimate and interpret directed and undirected edges in chain graphs under the presence of heavier-than-normal marginal tails. We incorporate the non-normality by proposing a random Gaussian-scale transformation of the original variables so that the transformed data is a Gaussian chain graph (GCM). The RCGM provides robust learning frameworks for various types of graphical models as special cases of chain graphs such as undirected networks by covariance/precision matrix specification and Bayesian networks when the entire topological order is known.

The increased scope of our model across heavy-tailed distributions comes at the expense of a Markov property weaker than conditional independence (CI), known as conditional sign-independence (CSI), to characterize dependencies for non-normal nodes. We assign each node a measure of its non-normality and use these scores to derive a probabilistic interpretation of CI and CSI properties. We incorporate sparsity in the chain graph by spike-and-slab priors on coefficients of the layer-wise regressions, and design the estimation algorithm to overcome the computational challenges that come with high-dimensional graphs using the node-wise likelihoods strategy \citep{ha2020bayesian}. Furthermore, we show that our algorithm outperforms GCM methods in terms of graph structural recovery under various degrees of non-normality exhibited in the datasets.

From a scientific perspective, heterogeneity in drug responses, even for standard of care for different cancers including lung, demands genomic-based drug treatments developed by integrating molecular and clinical data across several biological domains. We perform integrative network analysis using our RCGM algorithm on genomic, transcriptomic, proteomic and drug response data for lung cancer cell lines, available in the Cancer Dependency Map (\url{www.depmap.org}). We analyze the ways in which genomic features across key signaling pathways interact with each other and with mono-drug actions. From a global analysis of pathway-level networks, we identify pathways and genomic platforms most receptive to drugs. We find the DNA Damage Response (DDR) pathway to be the most connected with drugs and highlight its underlying dependencies.

In translational cancer research, diverse cancer models such as {\it in vivo} patient-derived xenografts (PDXs) have emerged as preclinical models that offer more faithful representation of genomic landscape of tumor and clinical outcomes than cancer cell lines \citep{gao2015high,woo2021conservation}. The growing number of PDX resources, e.g., PDXFinder, \url{pdxfinder.org} \citep{conte2019pdx} have facilitated systematic identification and validation of druggable genomic events. The holistic characterization of information flow of relevant mechanisms of drug sensitivity and resistance in various model systems of human cancer can further the development of new targeted therapies including combination treatments. Our integrative analysis framework is expected to aid in the identification of key molecular processes that drive clinical outcomes across different cancer types and populations, which can further help in developing genomic testing-based precision medicine. 
\bigskip
\begin{center}
{\large\bf SUPPLEMENTARY MATERIAL}
\end{center}
\bigskip
\vspace{-20pt}
The Supplementary Material contains proofs of Theorem~\ref{th: condl sign indep} and Corollary~\ref{th: chain graph condl sign indep 3}, details on the selection of mixing distributions, a description of MCMC sampling steps, and supplementary tables and figures. For reproducibility, the data and the R codes used for implementing our method are submitted with this paper.
\bigskip
\begin{center}
{\large\bf FUNDING}
\end{center}
\bigskip
\vspace{-20pt}
MJH was supported by the National Institutes	of	Health	grants R01CA244845-01A1 and R21CA22029, and start-up funds from University of Texas MD Anderson Cancer Center. VB was supported by the	National Institutes	of	Health grants R01-CA160736, R01CA244845-01A1, R21-CA220299, and P30 CA46592, US National Science Foundation grant 1463233, and start-up funds from the U-M Rogel Cancer Center and School of Public Health.
 AB was supported by US National Science Foundation Grant DMS-2014371.
\vspace{-15pt}
\bibliographystyle{apalike}
\bibliography{bibfileNonnormal}

\renewcommand{\theequation}{S.\arabic{equation}}
\renewcommand{\thesection}{S.\arabic{section}}  
\renewcommand{\thetable}{S.\arabic{table}}  
\renewcommand{\thefigure}{S.\arabic{figure}}
\setcounter{equation}{0}
\setcounter{section}{0}
\setcounter{figure}{0}
\setcounter{table}{0}
\newpage
\setcounter{page}{1}
\renewcommand{\thepage}{S.\arabic{page}}
\begin{center}
\singlespacing
    \Large{\textbf{Supplementary Materials for} \\ \emph{{Bayesian Robust Learning in Chain Graph Models for Integrative Pharmacogenomics}}}
\end{center}
\doublespacing
\date{}
\maketitle
\section{Proofs}\label{sec: supp proofs} This section consists of the proofs of Theorem~\ref{th: condl sign indep} and Corollary~\ref{th: chain graph condl sign indep 3}. We adapt the proof of Proposition 1 of \cite{bhadra2018inferring} to our chain graph context.
\subsection{Proof of Theorem~\ref{th: condl sign indep} (i)}\label{sec: supp. proof 2.1. part 1}
(a) Let $\mathcal{L}(u)=\mathcal{L}(v)=l$. Let $\mathrm{pa}_l$ be the indices of nodes in layers $1:(l-1)$. By construction of the non-normal chain graph model in \eqref{eq: nn-chain graph building}, $\boldsymbol X_l$ given $\boldsymbol D$ and $\boldsymbol X_{\mathrm{pa}_l}$ can be written as
\begin{align*}
     \boldsymbol D_l\boldsymbol X_l | \boldsymbol X_{\mathrm{pa}_l}, \boldsymbol D &\sim N(\boldsymbol B_{l, \mathrm{pa}_l}\boldsymbol D_{\mathrm{pa}_l}\boldsymbol X_{\mathrm{pa}_l}, \mathcal{K}_l^{-1}).
\end{align*}
Let $\boldsymbol g_l$ be equal to $\boldsymbol B_{l, \mathrm{pa}_l}\boldsymbol D_{\mathrm{pa}_l}\boldsymbol X_{\mathrm{pa}_l}$, so that $\mathrm{E}(\boldsymbol D_l\boldsymbol X_l | \boldsymbol X_{\mathrm{pa}_l}, \boldsymbol D) = \boldsymbol g_l$. Note that $\boldsymbol g_l$ is free of $\boldsymbol X_l$ and $\boldsymbol D_l$. Also, let $\boldsymbol g_l=((g))_j, j\in\mathcal{T}_l$. 
By properties of the multivariate normal distribution, 
\begin{align}\label{eq: general eq 1}
    \left(\frac{X_u}{d_u}, \frac{X_v}{d_v} \right)\big| \boldsymbol X_{[1:l]\backslash\{u, v\}}, \boldsymbol D \sim N_2\left( \left(\begin{array}{c}
         \mu_{u\boldsymbol D} \\
         \mu_{v\boldsymbol D}
    \end{array}\right), \mathcal{K}^{-1}_{\{u, v\}}
    \right),
\end{align}
where $\mu_{j\boldsymbol D}=E(X_j/d_j |\boldsymbol X_{[1:l]\backslash\{u, v\}},\boldsymbol D)$, $j=u, v$, and
$$ \mathcal{K}_{\{u, v\}} = \left(\begin{array}{cc}
    k_{uu} & k_{uv} \\
    k_{vu} & k_{vv}
\end{array}\right).$$
When $k_{uv}=0$, the joint conditional likelihood factorizes into the product of the individual conditional likelihoods, so that
\begin{align}\label{eq: general eq 2}
    f\left(\frac{X_u}{d_u}, \frac{X_v}{d_v} \bigg |\boldsymbol X_{[1:l]\backslash\{u,v\}},\boldsymbol D\right) &= \left( \frac{1}{\sqrt{2\pi}}\right)^2 k_{uu}^{1/2}\exp\left( -\frac{1}{2}\left(\frac{X_u}{d_u}-\mu_{u\boldsymbol D}\right)^Tk_{uu}\left(\frac{X_u}{d_u}-\mu_{u\boldsymbol D}\right) \right) \nonumber\\
    & \qquad \qquad k_{vv}^{1/2}\exp\left( -\frac{1}{2}\left(\frac{X_v}{d_v}-\mu_{v\boldsymbol D}\right)^Tk_{vv}\left(\frac{X_v}{d_v}-\mu_{v\boldsymbol D}\right) \right).
\end{align}
From Proposition C.5 of \cite{lauritzen1996graphical}, $\mu_{j\boldsymbol D}=E(X_j/d_j |\boldsymbol X_{[1:l]\backslash\{u, v\}},\boldsymbol D)$, $j=u,v$, can be deduced as
$$ \mu_{j\boldsymbol D} = g_j -\frac{1}{k_{jj}}\sum_{t\neq u, v;t\in \mathcal{T}_l} k_{jt}\left(\frac{X_t}{d_t}- g_t\right),\ \ j=u,v.$$
From \eqref{eq: general eq 2}, $(X_u/d_u)$ and $(X_v/d_v)$ are conditionally independent given $(\boldsymbol X_{[1:l]\backslash\{u,v\}},\boldsymbol D)$. Therefore, when $k_{uv}=0$, $(Y_j/d_j) |\boldsymbol X_{[1:l]\backslash\{u,v\}},\boldsymbol D \sim N(\mu_{j\boldsymbol D}, k_{jj}^{-1})$, $j=u,v$.
For the rest of the proof of part (i), we prove results involving $X_u$. The corresponding results for $X_v$ follow in an exact similar way. We evaluate 
\begin{align}\label{eq: sign prob 1}
    \mathrm{P}(X_u<0|\boldsymbol X_{[1:l]\backslash\{u,v\}},\boldsymbol D) &= \mathrm{P}\left(\frac{X_u}{d_u}<0 \bigg|\boldsymbol X_{[1:l]\backslash\{u,v\}},\boldsymbol D\right) \nonumber\\
    &= \mathrm{P}\left(k_{uu}^{1/2}\left(\frac{X_u}{d_u}-\mu_{u\boldsymbol D}\right)< -k_{uu}^{1/2} \mu_{u\boldsymbol D}\bigg |\boldsymbol X_{[1:l]\backslash\{u,v\}},\boldsymbol D\right) \nonumber\\
    &= \Phi(-k_{uu}^{1/2} \mu_{u\boldsymbol D}).
\end{align}
Note that since $\mu_{u\boldsymbol D}$ is free of $(X_u, X_v, d_u, d_v)$, the RHS of \eqref{eq: sign prob 1} is free of those as well. Similar calculation follows for $X_v$, so that $\mathrm{P}(X_v<0|\boldsymbol X_{[1:l]\backslash\{u,v\}},\boldsymbol D)=\Phi(-k_{vv}^{1/2} \mu_{v\boldsymbol D})$. 

Next, by properties of multivariate normal distribution, we have that
$$ (X_u/d_u) |\boldsymbol X_{[1:l]\backslash\{u\}}, \boldsymbol D \ \sim \ N(\tilde{\mu}_{u\boldsymbol D}, k_{uu}^{-1}),$$
where $$\tilde{\mu}_{u\boldsymbol D} = g_u -\frac{1}{k_{uu}}\sum_{t\neq u}k_{ut}\left(\frac{X_t}{d_t}- g_t\right).$$
Then it follows that
\begin{align}\label{eq: sign prob 2}
    \mathrm{P}(X_u<0|\boldsymbol X_{[1:l]\backslash\{u\}},\boldsymbol D) &= \mathrm{P}\left(\frac{X_u}{d_u}<0|\boldsymbol X_{[1:l]\backslash\{u\}},\boldsymbol D\right) \nonumber\\
    &= \mathrm{P}\left(k_{uu}^{1/2}\left(\frac{X_u}{d_u}-\tilde\mu_{u\boldsymbol D}\right)< -k_{uu}^{1/2} \tilde\mu_{u\boldsymbol D}|\boldsymbol X_{[1:l]\backslash\{u\}}, \boldsymbol D\right) \nonumber\\
    &= \Phi(-k_{uu}^{1/2} \tilde\mu_{u\boldsymbol D}).
\end{align}
When $k_{uv}=0$, the term corresponding to $t=v$ in the expression of $\tilde\mu_{u\boldsymbol D}$ vanishes, so that 
\begin{align*}
    \tilde\mu_{u\boldsymbol D} &= \mu_{u\boldsymbol D}.
\end{align*}
Therefore, when $k_{uv}=0$, \eqref{eq: sign prob 1} and \eqref{eq: sign prob 2} are equal, so that $\mathrm{P}(X_u<0|\boldsymbol X_{[1:l]\backslash\{u,v\}}, \boldsymbol D)$ is equal to $\mathrm{P}(X_u<0|\boldsymbol X_{[1:l]\backslash\{u\}},\boldsymbol D)$. 

Now note that the conditional likelihood of $\boldsymbol{DX}$ given $\boldsymbol D$ factorizes in terms of $d_v$'s, and $d_v\sim p_v$ are chosen independent of each other. Therefore $\pi(\boldsymbol D|\boldsymbol X)$ factorizes into the product of $\pi(1/d_v|X_v)$. Using this fact, we evaluate
\begin{align}\label{eq: condl sign indep undir}
    \mathrm{P}(X_u<0|\boldsymbol X_{[1:l]\backslash\{u,v\}}) &= \int \mathrm{P}(X_u<0|\boldsymbol X_{[1:l]\backslash\{u,v\}}, \boldsymbol D_{[1:l]}) d\pi(\boldsymbol D_{[1:l]}|\boldsymbol X_{[1:l]\backslash\{u,v\}}) \nonumber\\
    &= \int \mathrm{P}(X_u<0|\boldsymbol X_{[1:l]\backslash\{u\}}, \boldsymbol D_{[1:l]}) d\pi(\boldsymbol D_{[1:l]\backslash\{u,v\}}|\boldsymbol X_{[1:l]\backslash\{u,v\}}) \nonumber\\
    &= \int \mathrm{P}(X_u<0|\boldsymbol X_{[1:l]\backslash\{u\}}, \boldsymbol D_{[1:l]}) \left(\int d\pi(d_v^{-1}| \boldsymbol X_{[1:l]\backslash\{u,v\}})\right) \nonumber\\
    & \qquad\qquad\qquad\qquad\qquad\qquad d\pi(\boldsymbol D_{[1:l]\backslash\{u,v\}}|\boldsymbol X_{[1:l]\backslash\{u,v\}}) \nonumber\\
    &= \int \mathrm{P}(X_u<0|\boldsymbol X_{[1:l]\backslash\{u\}}, \boldsymbol D_{[1:l]}) d\pi(\boldsymbol D_{[1:l]\backslash u}|\boldsymbol X_{[1:l]\backslash\{u\}}) \nonumber\\
    &= \int \mathrm{P}(X_u<0|\boldsymbol X_{[1:l]\backslash\{u\}}, \boldsymbol D_{[1:l]}) d\pi(\boldsymbol D_{[1:l]}|\boldsymbol X_{[1:l]\backslash\{u\}}) \nonumber\\
    &=  \mathrm{P}(X_u<0|\boldsymbol X_{[1:l]\backslash\{u\}}).
\end{align}
The swapping of integrals is ensured by Fubini's theorem which can be applied here as $0<d_u<\infty$ and $\int p_u(d_u)<\infty$ by assumption on $d_v$s.

(b) Let $\mathcal{L}(u)<\mathcal{L}(v)$ and $\mathcal{L}(v)=l$. Note that $\boldsymbol X_{\{v\}\cup [1:l-1]}\boldsymbol D_{\{v\}\cup [1:l-1]}|\boldsymbol D$ follows a multivariate normal distribution. We then have, 
\begin{align*}
    \left(\frac{X_u}{d_u}, \frac{X_v}{d_v} \right)\big| \boldsymbol X_{[1:l-1]\backslash\{u\}},\boldsymbol D \sim N_2\left( \left(\begin{array}{c}
         \mu_{u\boldsymbol D} \\
         \mu_{v\boldsymbol D}
    \end{array}\right), \mathcal{K}^{-1}_{\{u, v\}}
    \right),
\end{align*}
where $\mu_{j\boldsymbol D}=E(X_j/d_j |\boldsymbol X_{[1:l-1]\backslash\{u\}},\boldsymbol D)$, $j=u, v$. Therefore, $$ (X_j/d_j) |\boldsymbol X_{[1:l-1]\backslash\{u\}},\boldsymbol D \ \sim \ N(\mu_{j\boldsymbol D}, k_{jj}^{-1}).$$
Also, for $j=u, v$, $$ (X_j/d_j) |\boldsymbol X_{[\{v\}\cup [1:l-1]]\backslash\{j\}},\boldsymbol D \ \sim \ N(\tilde{\mu}_{j\boldsymbol D}, k_{jj}^{-1}),$$ where $\tilde\mu_{j\boldsymbol D}=E(X_j/d_j |\boldsymbol X_{[\{v\}\cup [1:l-1]]\backslash\{j\}},\boldsymbol D)$.

Therefore, for $j=u, v$,
\begin{align}\label{eq: sign prob 1 dir}
    \mathrm{P}(X_j<0|\boldsymbol X_{[1:l-1]\backslash\{u\}},\boldsymbol D) &= \mathrm{P}\left(\frac{X_j}{d_j}<0|\boldsymbol X_{[1:l-1]\backslash\{u\}},\boldsymbol D\right) \nonumber\\
    &= \mathrm{P}\left(k_{jj}^{1/2}\left(\frac{X_j}{d_j}-\mu_{j\boldsymbol D}\right)< -k_{jj}^{1/2} \mu_{j\boldsymbol D}\bigg |\boldsymbol X_{[1:l-1]\backslash\{u\}},\boldsymbol D\right) \nonumber\\
    &= \Phi(-k_{jj}^{1/2} \mu_{j\boldsymbol D}).
\end{align}
Also,
\begin{align}\label{eq: sign prob 2 dir}
    \mathrm{P}(X_j<0|\boldsymbol X_{[\{v\}\cup [1:l-1]]\backslash\{j\}},\boldsymbol D) &= \mathrm{P}\left(\frac{X_j}{d_j}<0|\boldsymbol X_{[\{v\}\cup [1:l-1]]\backslash\{j\}},\boldsymbol D\right) \nonumber\\
    &= \mathrm{P}\left(k_{jj}^{1/2}\left(\frac{X_j}{d_j}-\tilde\mu_{j\boldsymbol D}\right)< -k_{jj}^{1/2} \tilde\mu_{j\boldsymbol D}|\boldsymbol X_{[\{v\}\cup [1:l-1]]\backslash\{j\}},\boldsymbol D\right) \nonumber\\
    &= \Phi(-k_{jj}^{1/2} \tilde\mu_{j\boldsymbol D}).
\end{align}
We now show that $\mu_{u\boldsymbol D}$=$\tilde{\mu}_{v\boldsymbol D}$ when $\boldsymbol B_{vu}=0$. To show this, note that
\begin{align*}
    \mathrm{E}(X_v/d_v|\boldsymbol X_{[1:l-1]},\boldsymbol D) &= \boldsymbol B _{v, [1:l-1]}\boldsymbol D_{[1:l-1]}\boldsymbol X_{[1:l-1]}\\
    &=\sum_{t\in [1:l-1]}\frac{\boldsymbol B_{vt}X_t}{d_t}\\
    &=\sum_{t\in [1:l-1]\backslash\{u\}}\frac{\boldsymbol B_{vt}X_t}{d_t}\\
    &=\mathrm{E}(X_v/d_v|\boldsymbol X_{[1:l-1]\backslash\{u\}},\boldsymbol D).
\end{align*}
Now note that since $\boldsymbol B_{vu}=0$, the construction of the chain graph (Equation~\eqref{eq: nn-chain graph building}) implies that $\boldsymbol B_{uv}=0$. Therefore it follows from the last display that $\mathrm{E}(X_u/d_u|\boldsymbol X_{[1:l-1]\backslash\{u\}},\boldsymbol D)=\mathrm{E}(X_u/d_u|\boldsymbol X_{\{v\}\cup [1:l-1]\backslash\{u\}},\boldsymbol D)$.

Therefore, when $B_{vu}=0$, $\tilde\mu_{j\boldsymbol D}=\mu_{j\boldsymbol D}$ for $j=u,v$, so that $\mathrm{P}(X_j<0|\boldsymbol X_{[1:l-1]\backslash\{u\}},\boldsymbol D)$ is equal to $\mathrm{P}(X_j<0|\boldsymbol X_{[\{v\}\cup [1:l-1]]\backslash\{j\}},\boldsymbol D)$. The rest of the proof involves integrating out $\boldsymbol D$ and follows similar to Equation~\eqref{eq: condl sign indep undir}.
\subsection{Proof of Theorem~\ref{th: condl sign indep} (ii)}\label{sec: supp. proof 2.1. part 2}
We prove the result for undirected edges. From \eqref{eq: general eq 2}, we have that $(X_u/d_u)$ and $(X_v/d_v)$ given $(\boldsymbol X_{[1:l]\backslash\{u,v\}}, \boldsymbol D)$ are jointly bivariate normal and are independent, so that  $$f(X_u/d_u|\boldsymbol X_{[1:l]\backslash\{u,v\}}, \boldsymbol D) = f(X_u/d_u|\boldsymbol X_{[1:l]\backslash\{u\}}, \boldsymbol D),$$
when $k_{uv}=0$. We then have
\begin{align}\label{eq: condl indep proof}
    f(X_u|\boldsymbol X_{[1:l]\backslash\{u,v\}}) 
    &= \int f(X_u|\boldsymbol X_{[1:l]\backslash\{u,v\}}, \boldsymbol D_{[1:l]}) d\pi(\boldsymbol D_{[1:l]\backslash u}|\boldsymbol X_{[1:l]\backslash\{u,v\}}) \nonumber\\
    &= \int d_u^{-1}f(X_u/d_u|\boldsymbol X_{[1:l]\backslash\{u,v\}}, \boldsymbol D_{[1:l]}) d\pi(\boldsymbol D_{[1:l]\backslash u}|\boldsymbol X_{[1:l]\backslash\{u,v\}}) \nonumber\\
    &= \int d_u^{-1}f(X_u/d_u|\boldsymbol X_{[1:l]\backslash\{u\}}, \boldsymbol D_{[1:l]}) d\pi(\boldsymbol D_{[1:l]\backslash u}|\boldsymbol X_{[1:l]\backslash\{u\}}) \nonumber\\
    &=\int f(X_u|\boldsymbol X_{[1:l]\backslash\{u\}}, \boldsymbol D_{[1:l]}) d\pi(\boldsymbol D_{[1:l]\backslash u}|\boldsymbol X_{[1:l]\backslash\{u\}}) \nonumber\\
    &= f(X_u|\boldsymbol X_{[1:l]\backslash\{u\}}). 
\end{align}
The interchange of integrals in $d\pi(\boldsymbol D)$ in the last display follows from arguments similar to \eqref{eq: condl sign indep undir}. The relation in Equation~\eqref{eq: general eq 2} also holds when $\mathcal{L}(u)<\mathcal{L}(v)$ and $\boldsymbol B_{vu}=0$, so the proof in this case would follow similar to the last two displays. 
\subsection{Proof of Corollary~\ref{th: chain graph condl sign indep 3}}\label{sec: supp. proof 2.2}
Let $\mathcal{L}(u)=\mathcal{L}(v)=l$. Let $H_1$ denote the event of conditional sign independence:
$$H_1 = \{\mathrm{P}(X_u<0|\boldsymbol X_{[1:l]\backslash\{u, v\}}) = \mathrm{P}(X_u<0|\boldsymbol X_{[1:l]\backslash u})\}.$$
Then $P(H_1)=\sum_{\omega_u,\omega_v}P(H_1|\omega_u,\omega_v)P(\omega_u,\omega_v)$, where $\omega_u,\omega_v$ are $0$ or $1$. From part (i) of Theorem~\ref{th: condl sign indep}, $H_1$ is true whenever at least one of $\omega_u,\omega_v$ is $1$. From part (ii) of Theorem~\ref{th: condl sign indep}, conditional independence is observed when $\omega_u=\omega_v=0$. As conditional sign independence is a weaker property, it is observed when $\omega_u=\omega_v=0$. So when $k_{uv}=0$,
\begin{align*}
   P(H_1|\omega_u=1,\omega_v=1)=P(H_1|\omega_u=1,\omega_v=0) &= P(H_1|\omega_u=0,\omega_v=1)\\
   &=P(H_1|\omega_u=0,\omega_v=0) = 1,
\end{align*}
so that $P(H_1)=1$.

For conditional independence, let $H_2$ denote the event:
$$H_2 = \{f(X_u|\boldsymbol X_{[1:l]\backslash\{u, v\}}) = f(X_u|\boldsymbol X_{[1:l]\backslash\{u\}})\}.$$
Then \begin{align*}
    P(H_2) &=\sum_{\omega_u,\omega_v}P(H_2|\omega_u,\omega_v)P(\omega_u,\omega_v)\\
    &\geq P(H_2|\omega_u=\omega_v=0)P(\omega_u=0,\omega_v=0).
\end{align*}
From part (ii) of Theorem~\ref{th: condl sign indep}, $P(H_2|\omega_u=\omega_v=0)=1$. As $\omega_v\sim \mathrm{Bern}(\pi_v)$ independently, $P(\omega_u=0,\omega_v=0)=1-\pi_u - \pi_v +\pi_u\pi_v$. Therefore $P(H_2)=1-\pi_u - \pi_v +\pi_u\pi_v$. Similar calculations would follow for $X_v$, and for $\mathcal{L}(u)<\mathcal{L}(v)$.
\section{Node-wise Likelihood Equations}\label{sec: prior formulation supp.}
 For a node $v$ belonging to layer $l$, let $\boldsymbol b_v$ be the entries in the row of $\boldsymbol B_l$ corresponding to $v$, with $\boldsymbol b_v=\boldsymbol 0$ if $l=1$. Let $\boldsymbol a_v=-k_{vv}^{-1}\boldsymbol k_l^{(v)}$, where $\boldsymbol k_l^{(v)}$ is the vector of $k_{vu}$, $u\in \mathcal{T}_l\backslash v$. We reparameterize the precision parameters to regression coefficients on residuals after taking out the effects of previous layers, so that given $\boldsymbol D$, we have 
 \begin{align*}
    (X_v/d_v) &= (\boldsymbol X\boldsymbol D)^T_{[1:\mathcal{L}(v)-1]}\boldsymbol b_v + \epsilon_v, 
 \end{align*} where $\boldsymbol\epsilon = (\epsilon_1,\ldots , \epsilon_q)$, $\epsilon_v=\boldsymbol\epsilon_{\mathcal{T}_l\backslash v}^T\boldsymbol a_v + e_v$, and $e_v\sim N(0, k_{vv}^{-1})$ is independent of $\boldsymbol \epsilon_{V\backslash v}$. Let $\boldsymbol B_l^{(v)}$ be the submatrix of $\boldsymbol B_l$ consisting of all but the row corresponding to the node $v$. Applying the node-wise regression in Proposition 1 in \cite{ha2020bayesian}, for the scaled random variables
given $\boldsymbol D$, we have the node-conditional equation
    \begin{align*}
        \frac{X_v}{d_v} &=(\boldsymbol X\boldsymbol D)^T_{[1:l-1]}(\boldsymbol b_v  - \boldsymbol B_l^{(v)}\boldsymbol a_v)+ (\boldsymbol X\boldsymbol D)^T_{\mathcal{T}_l\backslash v}\boldsymbol a_v + e_v.
    \end{align*}
With the identifiability constraints imposed, given $\boldsymbol D$, the $v$-th node-conditional regression is of the form
\begin{align*}
    (X_v/d_v) &= (\boldsymbol X\boldsymbol D)^T_{[1:l-1]}\boldsymbol b_v + \boldsymbol \epsilon_{\mathcal{T}_l\backslash v}^T\boldsymbol a_v  + e_v,
\end{align*}where $\epsilon_{\mathcal{T}_l\backslash v}^T=(\boldsymbol X\boldsymbol D)^T_{\mathcal{T}_l\backslash v} - (\boldsymbol X\boldsymbol D)^T_{[1:l-1]}\boldsymbol B_l^{(v)}$ is assumed fixed given $\boldsymbol D$. We reparameterize $\boldsymbol a_v$, $\boldsymbol b_v$, $k_{vv}$ for every $v\in V$ and $\mathcal{L}(v)=l$, to obtain the layer-wise estimates of $\boldsymbol B$ and $\mathcal{K}$ as
\begin{align}\label{eq: reparametrize back B K supp}
    ((\boldsymbol B_l))_{vu} &= \boldsymbol b_{vu},\nonumber\\
    ((\mathcal{K}_l))_{vu} &= -k_{vv} a_{vu}, \quad u\in \mathcal{T}_l\backslash v.
\end{align}


\section{Identifiability in Node-wise Likelihood Equation}\label{sec: identifiability}
\begin{figure}[h!]
    \centering
    \includegraphics[width=0.85\textwidth]{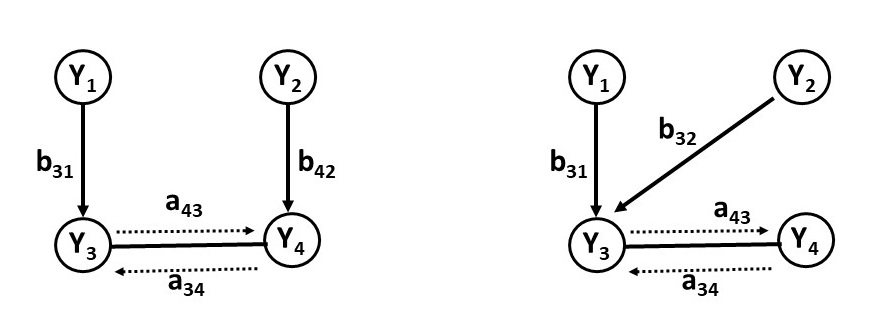}
    \caption{An example demonstrating the node-wise regressions for a two-layer chain graph with two nodes in each layer}
    \label{fig: chaingraph toy eg}
\end{figure}

Throughout this section, let $\boldsymbol Y=\boldsymbol D\boldsymbol X$ denote the scaled data. Let us examine a simple case of the model as described in Figure~\ref{fig: chaingraph toy eg}. The node-wise equation for $Y_3$ would be
\begin{align*}
    Y_3 &= (b_{31}Y_1 + b_{32}Y_2) + a_{34}Y_4 - a_{34}b_{42}Y_2.
\end{align*}
From the figure, it appears that $b_{32}=0$, but $b_{32}$ would still appear in the node-wise regression equation. The previous equation can be re-written as
\begin{align*}
    Y_3 &= b_{31}Y_1 + (b_{32} - a_{34}b_{42})Y_2 + a_{34}Y_4.
\end{align*}
Now let $\hat{b}$ be the estimated coefficient for $Y_2$. We assume $a_{34}$ known when estimating $b$, so $\hat{b}=b_{32}-a_{34}b_{42}$ is one equation with two variables $b_{32}$ and $b_{42}$ which cannot be solved uniquely. As an example, consider the two different sets of solutions:
\begin{enumerate}
    \item $(b_{32}=0, b_{42}=-\hat{b}/a_{34})$: graph on the left panel. There is no edge between $Y_3$ and $Y_2$.
    \item $(b_{32}=\hat{b}, b_{42}=0)$: graph on the right panel. There is no edge between $Y_4$ and $Y_2$.
\end{enumerate}
In this example, it is not possible to identify whether the effect of $Y_2$ on $Y_3$ is through the directed edge $b_{32}$ or via the indirect effect of $Y_4$ (through $a_{34}b_{42}$).
So the identifiability problem can potentially lead to inconsistent edge detection. 
\section{Selection of Mixing Distributions}\label{sec: selection of mixing dist}
If $v$ is a node with heavy tails such that $\omega_v=1$, then 
\begin{align}
    d_v|(\omega_v=1) &\sim  p_v,\nonumber\\
    f(X_v|\omega_v=1) & =  \int d_v^{-1}\phi(X_v/d_v)dp_v(d_v)  ,
\end{align}
where $\phi(\cdot)$ is the pdf of a standard normal distribution. From Lemma 1 of \cite{bhadra2018inferring}, if the tail of $\boldsymbol X_v$ decays polynomially as $\alpha|x|^{2\lambda_v - 1}$ for some $\lambda\leq 0$, $\nu\in {\rm I\!R}$, $\alpha\in (0, 1)$ then the mixing density $p_v$ has the right-tail decaying as $\alpha d_v^{\lambda_v - 1}$. Further, if the marginal decays exponentially as $|x|^{2\lambda_v - 1}\exp\{-(2\psi_v)^{1/2}|x_v|\}$ for $\lambda\in {\rm I\!R}$ and $\psi >0$, then $p_v$ has tail decaying as $d_v^{\lambda_v - 1}\exp(-\psi_vd_v)$. So the rate of decay in marginal tails can be directly used to set the parameters of the mixing distribution $p_v$. Algorithm~\ref{alg: alg 1} describes the procedure for setting the tail parameters of $p_v$.

\begin{algorithm}[t]
\SetAlgoLined
 Given the multivariate data $\boldsymbol X$ with $q$ coordinates,\;
 \While{$1\leq i\leq q$}{
  Let $\boldsymbol x=[\boldsymbol X_v - \text{mean}(\boldsymbol X_v)]/\text{s.d.}\boldsymbol (X_v)$\;
  (\textbf{double-exponential fitting}). Evaluate $p_e$, the \emph{p}-value of the regression $\log\hat{f}(x)=a_0 + a_1\log |x| + a_2|x|$.\;
  (\textbf{t-distribution fitting}). Evaluate $p_t$, the \emph{p}-value of the regression $\log\hat{f}(x)=a_0 + a_2|x|$.\;
  (\textbf{select category and parameters of $p_v$}). \If{$p_e<p_t$}{
  Set $p_v$ as Gamma(shape$=(a_1+1)/2$, scale$=a_1^2/2$)}
  \Else{Set $p_v$ as Inverse-Gamma(shape$=(-a_1+1)/2$, scale$=(-a_1+1)/2$)}
 }
 \caption{ Selection of mixing distribution $p_v$'s}
 \label{alg: alg 1}
\end{algorithm}
\section{MCMC Sampling Steps Summary}\label{supp:mcmc}
We first choose $p_v$ for each node $v$ using Algorithm~\ref{alg: alg 1}. We then set the prior of $\pi_v$ as $\text{Beta}(\mu_vr_v, (1-\mu_v)r_v)$, where $\mu_v=H(\boldsymbol x_v)$, $H(\boldsymbol x_v) = 2*\Phi(\log(1-pval(\boldsymbol x_v)))$, with $\Phi$ and $pval$ respectively being the cdf of standard normal distribution and the \emph{p}-value of the Kolmogorov-Smirnov test for normality, $r_v=[\mu_v(1-\mu_v)/\xi_v]-1$, with the prior variance $\xi_v$ set to $0.01$ if $s_v>0.01$ and $0.01s_v$ otherwise, for $s_v=\mu_v(1-\mu_v)$. We center and scale each coordinate of $\boldsymbol X$ before computation. At iteration $t$ of the MCMC sampling, we generate $\boldsymbol D$ for every subject $i$ given current $\boldsymbol \pi$ as $d_{iv}\sim \pi_vp_v + (1-\pi_v)\delta_1$ for all $v\in V$. We then accept the sampled $d_{iv}$ based on the ratio
\begin{align}\label{eq: MCMC generate di}
    R &= \frac{\phi(X_{iv}d_{iv})p_v(d_{iv}^*)}{\phi(X_{iv}d_{iv}^*)p_v(d_{iv})},
\end{align}
where $d_{iv}^*$ is the current value of $d_{iv}$ and $\phi$ is the density of a standard normal distribution. We accept $d_{iv}$ if $U\sim\text{Unif}(0,1)$ is less than $R$. Next, for every layer $l$, we update the normality measures and directed and undirected edges as described below.

To update the non-normality measure $\boldsymbol \pi$, let $\boldsymbol\pi^s$ be the current state of $\boldsymbol\pi$. We generate the new sample $\boldsymbol\pi^*$ by generating $\pi_v^*$ for every $v\in \mathcal{T}_l$ from $\text{Beta}(\xi_v^0(\xi_v^0/r_v^2-1), (1-\xi_v^0)(\xi_v^0/r_v^2-1))$. For every $v\in\mathcal{T}_l$, we look at the acceptance ratio

\begin{align}\label{eq: MCMC update pi}
    R &= \frac{[\pi_v^* \phi(\tilde{\boldsymbol X}_v/d_v|\mu_v^D, \boldsymbol\pi^*,  k_{vv})p_v(d_v) + (1-\pi_v^*)\phi(\tilde{\boldsymbol X}_v|\mu_v, \boldsymbol\pi^*,  k_{vv})]g(\pi_v^*)}{[\pi_v^s \phi(\tilde{\boldsymbol X}_v/d_v|\mu_v^D, \boldsymbol\pi^*,  k_{vv})p_v(d_v) + (1-\pi_v^s)\phi(\tilde{\boldsymbol X}_v|\mu_v, \boldsymbol\pi^*,  k_{vv})]g(\pi_v^s)},
\end{align}
 where $g$ is the density function of $\text{Beta}(\xi_v^0(\xi_v^0/r_v^2-1), (1-\xi_v^0)(\xi_v^0/r_v^2-1))$, and $\mu_v= \boldsymbol X_{P_v}^T\boldsymbol b_v - \boldsymbol {X}_{\mathcal{T}_l\backslash v}\boldsymbol a_v + \boldsymbol{X_{P_v}B}_{\mathcal{T}_l\backslash v}^T\boldsymbol a_v$, $\mu_v^{\boldsymbol D}= \boldsymbol X_{P_v}^T\boldsymbol b_v - (\boldsymbol {XD})_{\mathcal{T}_l\backslash v}\boldsymbol a_v + \boldsymbol{X_{P_v}B}_{\mathcal{T}_l\backslash v}^T\boldsymbol a_v$.
 We then sample $U\sim\text{Unif}(0, 1)$ and set $\boldsymbol\pi^{s+1}=\boldsymbol\pi^*$ if $U\le R$, and set $\boldsymbol\pi^{s+1}=\boldsymbol\pi^s$ otherwise.
 
To update undirected edges, given $\boldsymbol D$, let $\tilde{\boldsymbol{y}}_v = \boldsymbol X_v/d_v - \boldsymbol X_{P_v} \boldsymbol b_v$, and $\boldsymbol x_v = \boldsymbol X_{\mathcal{T}_l\backslash v}D_{\mathcal{T}_l\backslash v} - \boldsymbol X_{P_v} \boldsymbol B^T_{\mathcal{T}_l\backslash v}$.
\begin{enumerate}
    \item We first update the model selection parameters. Let $s$ be the current state.
    \begin{enumerate}
        \item Add-delete or swap: with probability 1/2, sample $w_1$ from $\mathcal{T}_l\backslash v$ and set $\eta^*_{vw_1}=\eta^*_{w_1v}=1-\eta^s_{vw_1}$. Else, sample $w_2$ from $\{w: \eta_{vw}=0\}\backslash\{v\}$, $w_3$ from $\{w: \eta_{vw}=1\}$ and set $\eta_{vw_2}^*=\eta_{w_2v}=1$ and $\eta_{vw_3}=\eta_{w_3v}=0$.
        \item Compute the acceptance ratio:\\
            add/delete:  $R = \prod_{r\in\{v, w_1\}} \frac{f(\tilde{\boldsymbol y}_r|\boldsymbol x_r, \boldsymbol\eta_r^*,  k_{rr})p(\boldsymbol\eta^*_r)}{f(\tilde{\boldsymbol y}_r|\boldsymbol x_r, \boldsymbol\eta_r^s,  k_{rr})p(\boldsymbol\eta^s_r)}$;\\
             swap:  $R = \prod_{r\in\{v, w_2, w_3\}} \frac{f(\tilde{\boldsymbol y}_r|\boldsymbol x_r, \boldsymbol\eta_r^*,  k_{rr})p(\boldsymbol\eta^*_r)}{f(\tilde{\boldsymbol y}_r|\boldsymbol x_r, \boldsymbol\eta_r^s,  k_{rr})p(\boldsymbol\eta^s_r)}$,\\
        where $f(\tilde{\boldsymbol y}_v|\boldsymbol x_v, \boldsymbol\eta_v,  k_{vv})$ is the density function of a normal distribution with mean 0 and variance 
        \begin{align*}
            \frac{1}{ k_{vv}}\left( \boldsymbol I - \boldsymbol x^\eta_v (\boldsymbol x^{\eta T}_v\boldsymbol x^\eta_v + \boldsymbol G_v^{-1})^{-1} \boldsymbol x^{\eta T}_v \right)^{-1},
        \end{align*}
        for $\boldsymbol G_v=\boldsymbol I/\lambda_l$ and $p(\boldsymbol\eta_r)=\prod_{k\in C(r)}p_{rk}^{\eta_rk}(1-p_{rk})^{1-\eta_rk}$.
        \item Sample $U\sim\text{Unif}(0, 1)$ and set $\boldsymbol\varepsilon^{s+1}=\boldsymbol\varepsilon^*$ if $U\le R$, and set $\boldsymbol\varepsilon^{s+1}=\boldsymbol\varepsilon^s$ otherwise.
    \end{enumerate}
    \item Gibbs sampling for $\boldsymbol  a_v$: for all $r$ in $\{v, w_1\}$ or $\{v, w_2, w_3\}$,
    \begin{align}\label{eq: sampling gibbs alpha 1}
        \boldsymbol a_r|\tilde{\boldsymbol y}_r, \tilde{\boldsymbol x}_r, \boldsymbol\eta_r,  k_{rr} &\sim N\left((\boldsymbol x^{\eta T}_r\boldsymbol x^\eta_r + \boldsymbol G_r^{-1})^{-1}\boldsymbol x^{\eta T}_r\tilde{\boldsymbol y}_r, \frac{1}{ k_{rr}}(\boldsymbol x^{\eta T}_r\boldsymbol x^\eta_r + \boldsymbol G_r^{-1})^{-1} \right).
    \end{align}
    \item Gibbs sampling for $ k_{rr}$: for all $r$ in $\{v, w_1\}$ or $\{v, w_2, w_3\}$,
    \begin{align}\label{eq: sampling gibbs kappa 1}
        p( k_{rr}|\tilde{\boldsymbol x}_r, \boldsymbol x_r, \boldsymbol\eta_r, \boldsymbol a_r) &= \text{Gamma}(A_1, A_2),
    \end{align}
    where 
    \begin{align*}
        A_1 &= \frac{n+\delta_l+|\mathcal{T}_l|-1+\|\boldsymbol\eta_r\|_0}{2}, \\
        A_2 &= \frac{\lambda_l}{2} + \frac{1}{2}\left[ (\tilde{\boldsymbol y}_r - \boldsymbol x_r^\eta\boldsymbol a_r^\eta)^T(\tilde{\boldsymbol y}_r - \boldsymbol x_r^\eta\boldsymbol a_r^\eta) + \boldsymbol b_r^{\gamma T}(\boldsymbol I/\lambda_l)^{-1}\boldsymbol b_r^\gamma \right.\\
        & \left. \qquad\qquad\qquad\qquad\qquad\qquad + (\lambda_l+|\mathcal{T}_l|-1)\boldsymbol a_r^{\eta T}\boldsymbol a_r^\eta\right], 
    \end{align*}
    and $\|\cdot\|_0$ denotes the number of nonzero elements in the vector.
\end{enumerate}

To update directed edges between layers, let $G$ be the current graph and $u_1, u_2, \ldots$ be vertices in the neighbor set of $v$ in $G$. Set $\tilde{\boldsymbol y}_v=X_v/d_v - \boldsymbol \epsilon_{\mathcal{T}_l\backslash v}^T\boldsymbol a_v$, $\tilde{\boldsymbol x}_v = \boldsymbol D_{[1:l-1]}\boldsymbol X_{[1:l-1]}$.
\begin{enumerate}
    \item Metropolis-Hastings for edge selection: let $s$ be the current state.
    \begin{enumerate}
        \item Add-delete or swap: with probability 1/2, sample $k_1$ from $\boldsymbol X_{[1:l-1]}$ and set $\gamma^*_{vk_1}=1-\gamma^s_{vk_1}$. Otherwise sample $k_2$ from the parent nodes of $v$, $k_3$ from the set of nodes in $[1:l-1]$ that are not connected to $v$, and set $\gamma_{vk_2}^*=0$ and $\gamma_{vk_3}=1$.
        \item Compute the acceptance ratio:
        \begin{align*}
            R &=  \frac{f(\tilde{\boldsymbol y}_v|\tilde{\boldsymbol x}_v, \boldsymbol\gamma^*, k_{vv})p(\boldsymbol\gamma^*)}{f(\tilde{\boldsymbol y}_v|\tilde{\boldsymbol x}_v, \boldsymbol\gamma^s, k_{vv})p(\boldsymbol\gamma^s)},
        \end{align*}
        where $f(\tilde{\boldsymbol y}_v|\tilde{\boldsymbol x}_v, \boldsymbol\gamma, k_{vv})$ is the density function of normal distribution with mean 0 and variance 
        \begin{align*}
            \frac{1}{ k_{vv}}\left( \boldsymbol I -  \boldsymbol x^\gamma_v ( \boldsymbol x^{\gamma T}_v\boldsymbol x^\gamma_v + \boldsymbol G_v^{-1})^{-1} \boldsymbol x^{\gamma}_v \right)^{-1},
        \end{align*}
        for $\boldsymbol G_v=\boldsymbol I/\lambda_l$ and $p(\boldsymbol\gamma)=\prod_{v\in \mathcal{T}_l, w\in\mathcal{T}_{[1:l-1]}}q_{vw}^{\gamma_{vw}}(1-q_{vw})^{1-\gamma_{vw}}$.
        \item Sample $U\sim\text{Unif}(0, 1)$ and set $\boldsymbol\gamma^{s+1}=\boldsymbol\gamma^*$ if $U\le R$, and set $\boldsymbol\gamma^{s+1}=\boldsymbol\gamma^s$ otherwise.
    \end{enumerate}
    \item Gibbs sampling for $\boldsymbol b_v$:
    \begin{align}\label{eq: sampling gibbs alpha 2}
        \boldsymbol b_v|\tilde{\boldsymbol y}_v, \tilde{\boldsymbol x}_v, \boldsymbol\gamma_v,  k_{vv} &\sim N\left((\boldsymbol x^{\boldsymbol\gamma T}_v\boldsymbol x^{\boldsymbol\gamma}_v + \boldsymbol G_v^{-1})^{-1}\boldsymbol x^{\boldsymbol\gamma T}_v\tilde{\boldsymbol y}_v, \frac{1}{ k_{vv}}(\boldsymbol x^{\boldsymbol\gamma T}_v\boldsymbol x^{\boldsymbol\gamma}_v + \boldsymbol G_v^{-1})^{-1} \right).
    \end{align}
    \item Gibbs sampling for $ k_{vv}$:
    \begin{align}\label{eq: sampling gibbs kappa 2}
        p( k_{vv}|\tilde{\boldsymbol y}_v, \tilde{\boldsymbol x}_v, \boldsymbol\gamma_v,  \boldsymbol b_v) &= \text{Gamma}(B_1, B_2),
    \end{align}
    where 
    \begin{align*}
        B_1 &= \frac{n+\delta_l+|\mathcal{T}_l|-1+\|\boldsymbol\gamma_v\|_0}{2}, \\
        B_2 &= \frac{\lambda_l}{2} + \frac{1}{2}\left[ (\tilde{\boldsymbol y}_v - \boldsymbol x_v^{\gamma T}\boldsymbol b_v^\gamma)^T(\tilde{\boldsymbol y}_v - \boldsymbol x_v^{\gamma T}\boldsymbol b_v^\gamma) + \boldsymbol b_v^{\gamma T}(\boldsymbol I/\lambda_l)^{-1}\boldsymbol b_v^\gamma \right.\\
        & \left. \qquad\qquad\qquad\qquad\qquad\qquad + (\lambda_l+|\mathcal{T}_l|-1)\boldsymbol b_v^{\gamma T}\boldsymbol b_v^\gamma\right], 
    \end{align*}
    and $\|\cdot\|_0$ is the number of nonzero elements in the vector.
\end{enumerate}
\section{Additional Simulation Details}\label{sec: supplementary simulation details}
\begin{figure}[!h]
 \includegraphics[width=\textwidth]{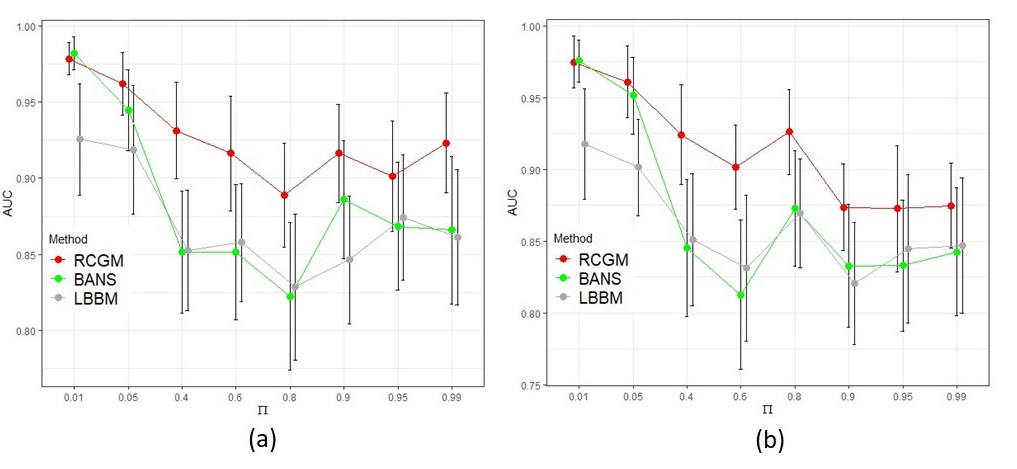}
     \caption{Mean area under ROC curve across $30$ replications for different levels of marginal tail-heaviness, measured by $\pi$. Interval of one standard deviation around each mean AUC is displayed through vertical lines. Panels (a) and (b) correspond to scaling by $\text{Exponential}(\text{mean}=2.5)$ and $\text{Inv-Gamma}(\text{shape}=3,\text{rate}=6)$ respectively. }
     \label{fig:AUCvsPi}
 \end{figure}

\begin{figure}[!h]
 \includegraphics[width=\textwidth]{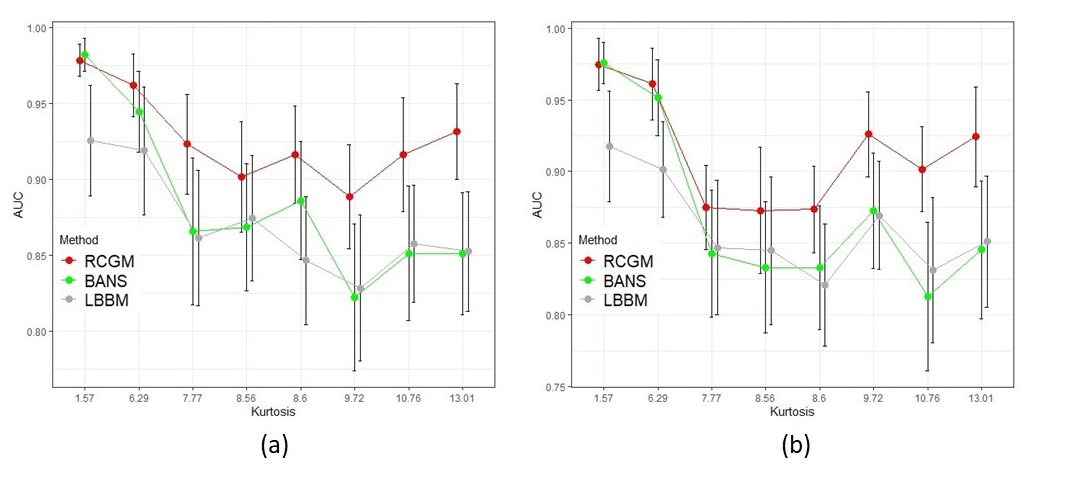}
     \caption{Mean area under ROC curve across $30$ replications for different levels of marginal tail-heaviness, measured by the kurtosis. Interval of one standard deviation around each mean AUC is displayed through vertical lines. Panels (a) and (b) correspond to scaling by $\text{Exponential}(\text{mean}=2.5)$ and $\text{Inv-Gamma}(\text{shape}=3,\text{rate}=6)$ respectively.}
     \label{fig:AUCvsPi0}
 \end{figure}

The most commonly used measure for the tail mass being heavier or lighter than that of the normal distribution is the kurtosis. A standardized kurtosis $>0$ indicates heavier-than-normal tails. To analyze the performance of the three methods with respect to tail-heaviness measured by the kurtosis, we empirically calculate the kurtosis of a random variable following a mixture of normal and heavy-tailed distributions with the mixing factor being $\pi$. We perform this procedure for a range of $\pi$ in $[0,1]$. Based on these $\pi$ values, we form the chain graph simulation datasets as described in Section~\ref{sec: simulations} and evaluate the AUC values for $30$ replications of data, for each of the three methods RCGM, BANS and LBBM. Figures~\ref{fig:AUCvsPi} -- \ref{fig:AUCvsPi0} display the AUCs varying across different $\pi$ and kurtosis levels respectively.
 \section{Pathway-wise Networks for Molecular-drug Interactions}\label{sec: supp pathway networks}
\begin{figure}
     \centering
     \includegraphics[scale=0.8]{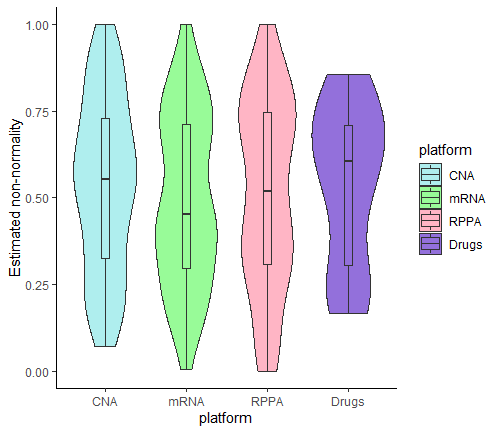}
     \caption{Violin plot of the estimated node-wise non-normality $\hat{\pi}_v$ across all pathways.}
     \label{fig:Estimated pi violin plot}
 \end{figure}

\begin{figure}
     \centering
     \includegraphics[scale=0.55]{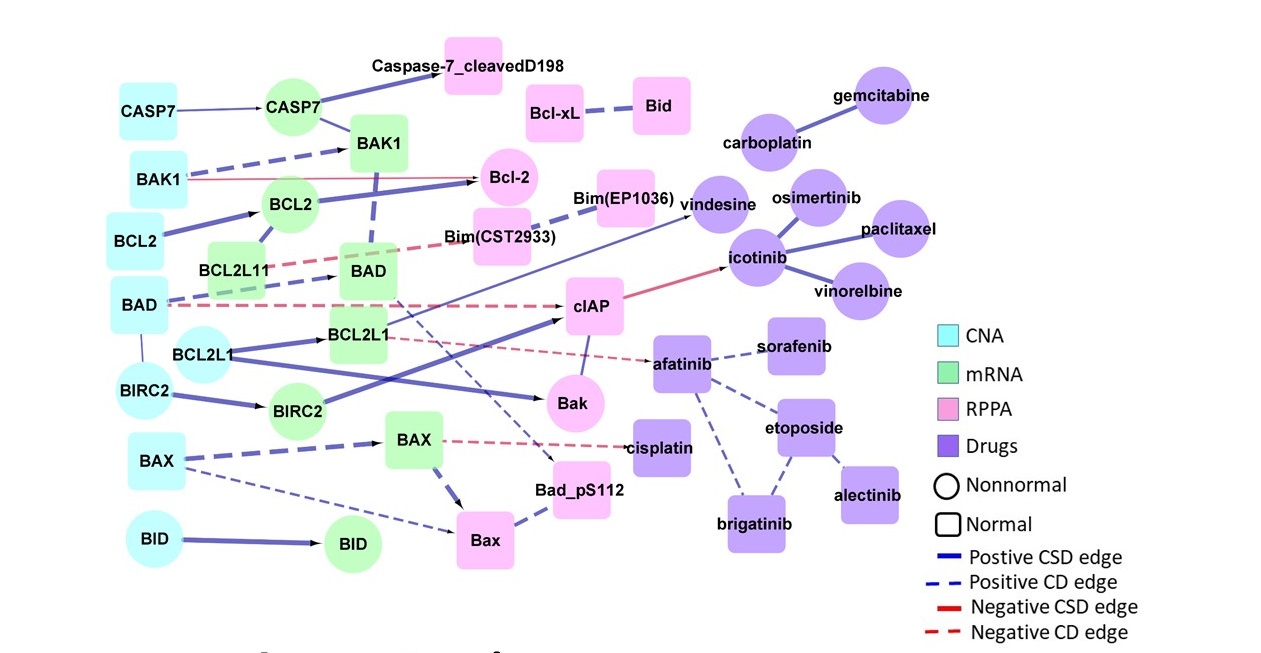}
     \caption{The estimated multilayered network for the apoptosis pathway. Blue and red edges indicate positive and negative dependencies, while CD and CSD stand for conditionally dependent and conditionally sign-dependent edges respectively. The width of the edges is proportional to the posterior inclusion probabilities.}
     \label{fig:graph_apoptosis}
 \end{figure}

\begin{figure}
     \centering
     \includegraphics[scale=0.55]{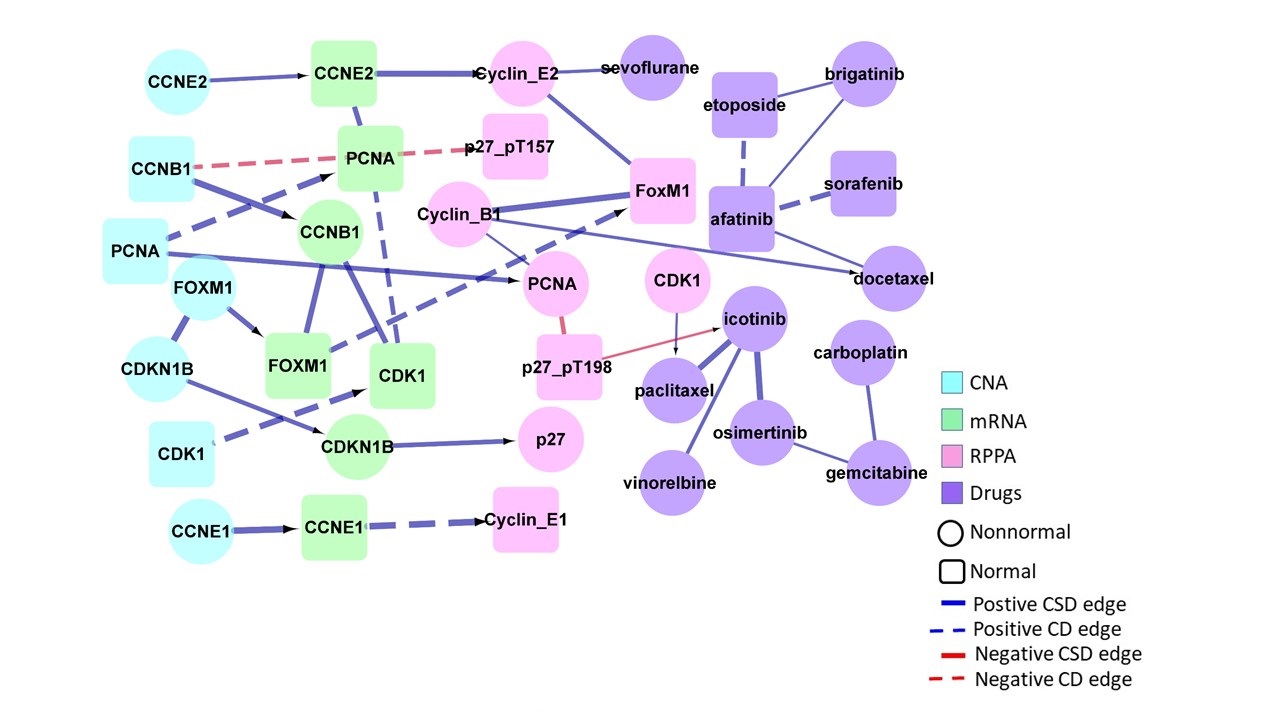}
     \caption{The estimated multilayered network for the cell cycle pathway. Blue and red edges indicate positive and negative dependencies, while CD and CSD stand for conditionally dependent and conditionally sign-dependent edges respectively. The width of the edges is proportional to the posterior inclusion probabilities.}
     \label{fig:graph_cellcycle}
 \end{figure}

\begin{figure}
     \centering
     \includegraphics[scale=0.55]{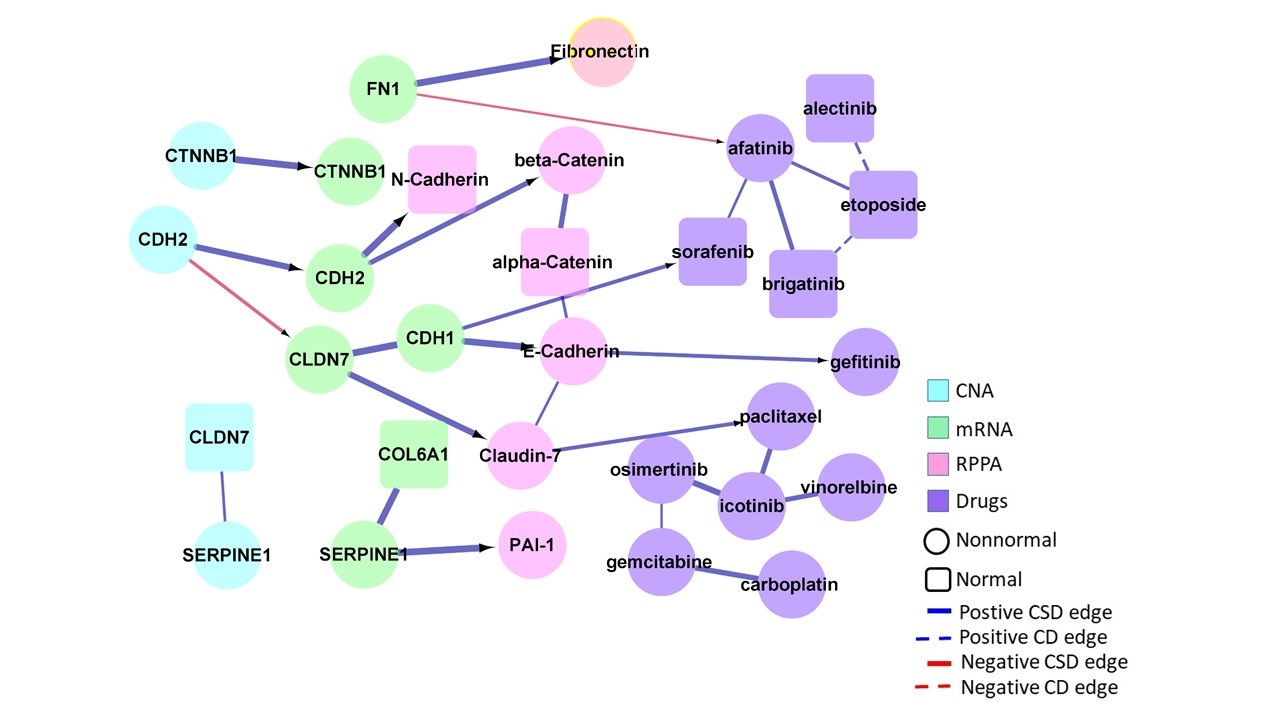}
     \caption{The estimated multilayered network for the EMT pathway. Blue and red edges indicate positive and negative dependencies, while CD and CSD stand for conditionally dependent and conditionally sign-dependent edges respectively. The width of the edges is proportional to the posterior inclusion probabilities.}
     \label{fig:graph_emt}
 \end{figure}

\begin{figure}
     \centering
     \includegraphics[scale=0.55]{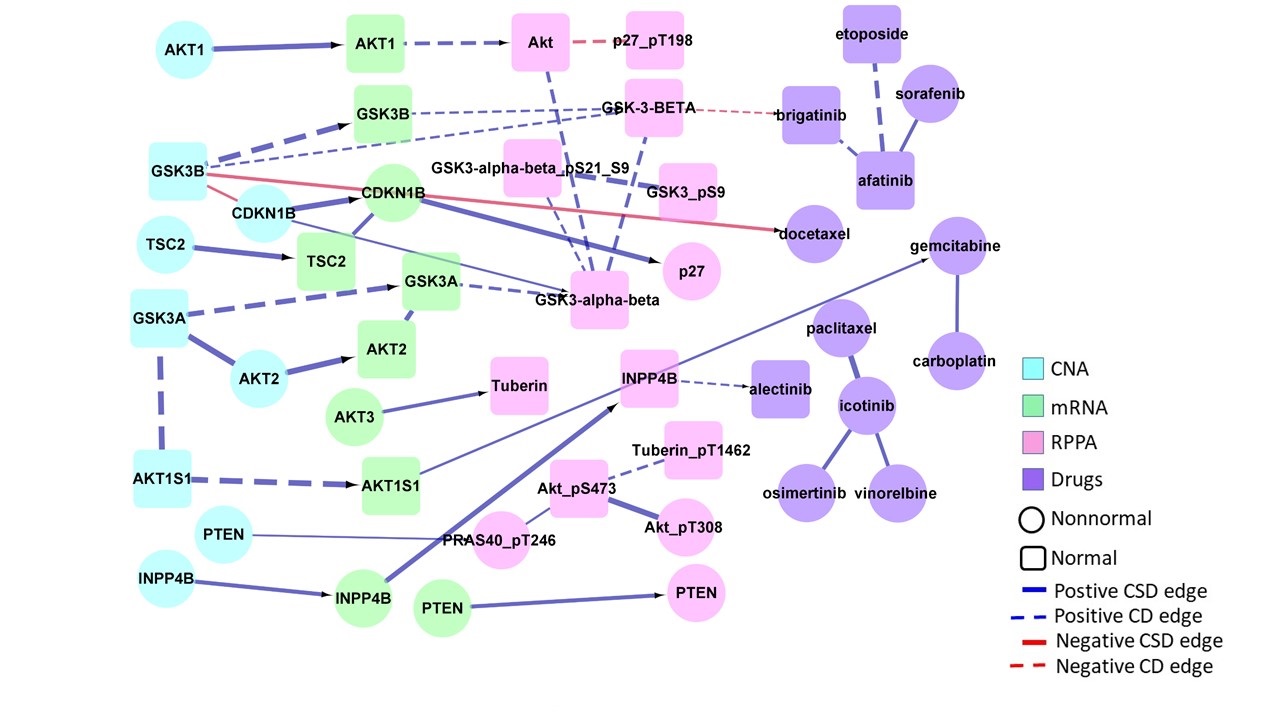}
     \caption{The estimated multilayered network for the PI3K/AKT pathway. Blue and red edges indicate positive and negative dependencies, while CD and CSD stand for conditionally dependent and conditionally sign-dependent edges respectively. The width of the edges is proportional to the posterior inclusion probabilities.}
     \label{fig:graph_pi3k}
 \end{figure}

\begin{figure}
     \centering
     \includegraphics[scale=0.55]{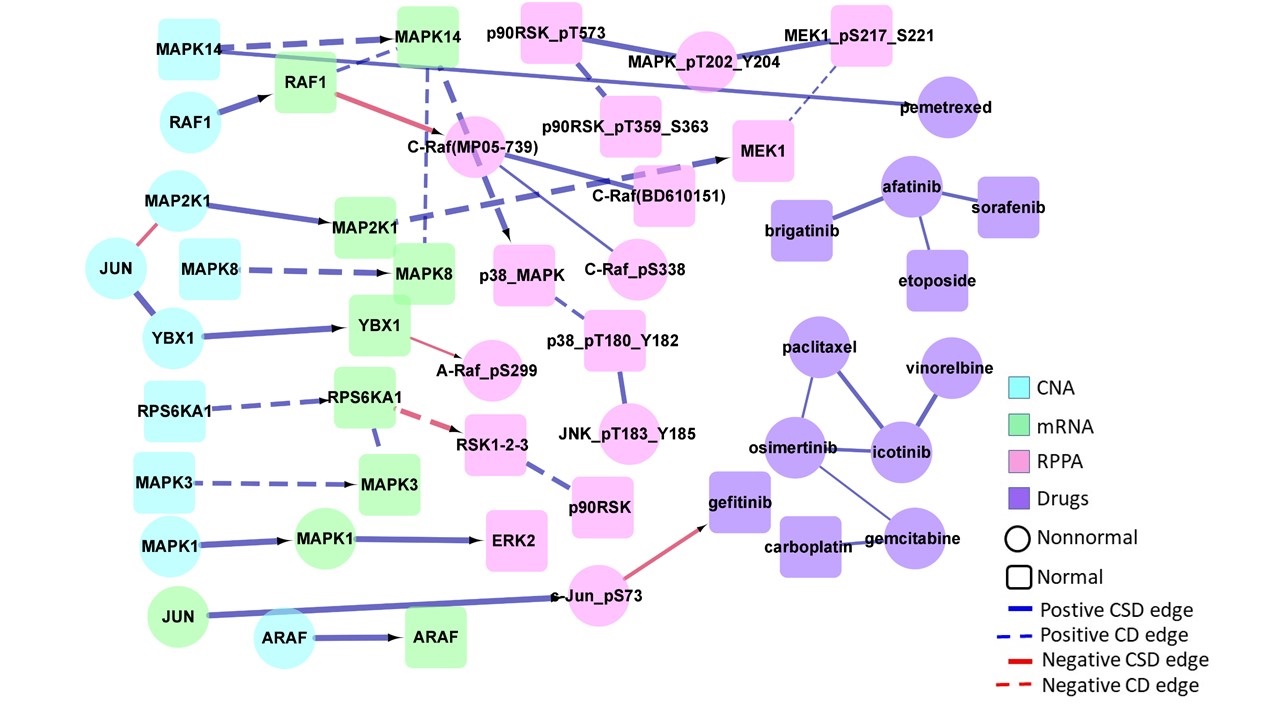}
     \caption{The estimated multilayered network for the RAS/MAPK pathway. Blue and red edges indicate positive and negative dependencies, while CD and CSD stand for conditionally dependent and conditionally sign-dependent edges respectively. The width of the edges is proportional to the posterior inclusion probabilities.}
     \label{fig:graph_rasmapk}
 \end{figure}

\begin{figure}
     \centering
     \includegraphics[scale=0.55]{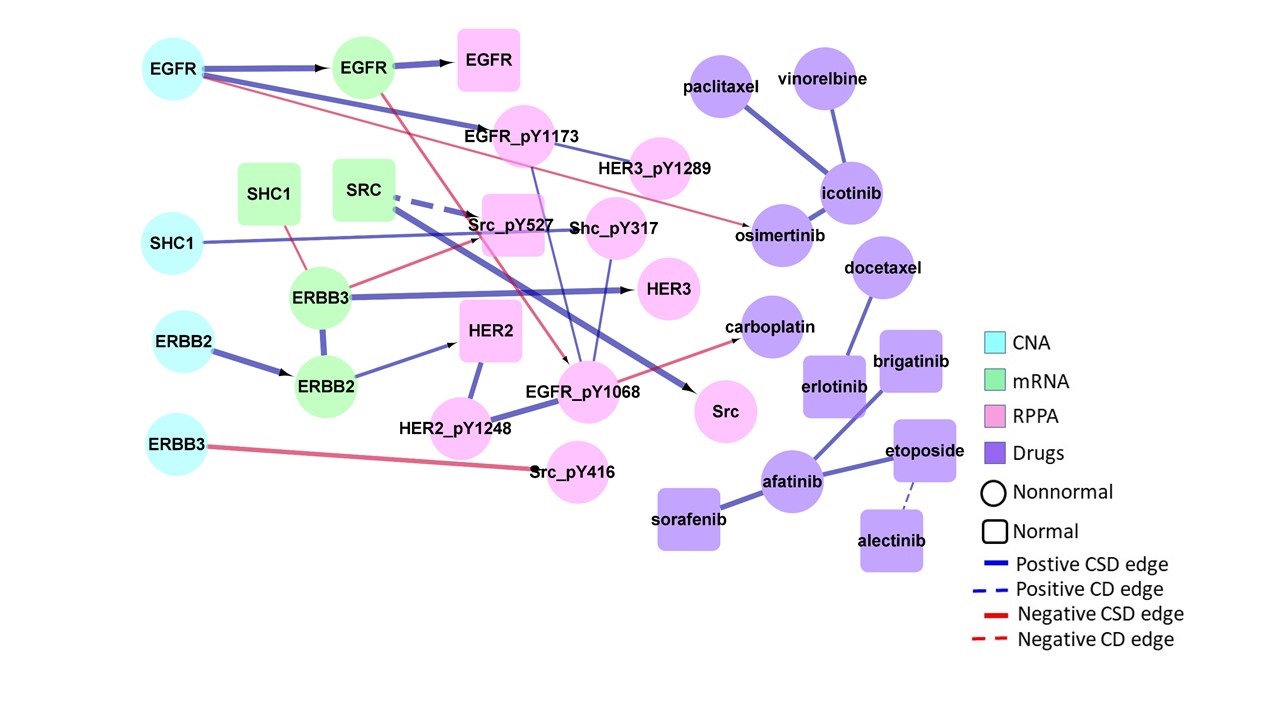}
     \caption{The estimated multilayered network for the RTK pathway. Blue and red edges indicate positive and negative dependencies, while CD and CSD stand for conditionally dependent and conditionally sign-dependent edges respectively. The width of the edges is proportional to the posterior inclusion probabilities.}
     \label{fig:graph_rtk}
 \end{figure}

\begin{figure}
     \centering
     \includegraphics[scale=0.55]{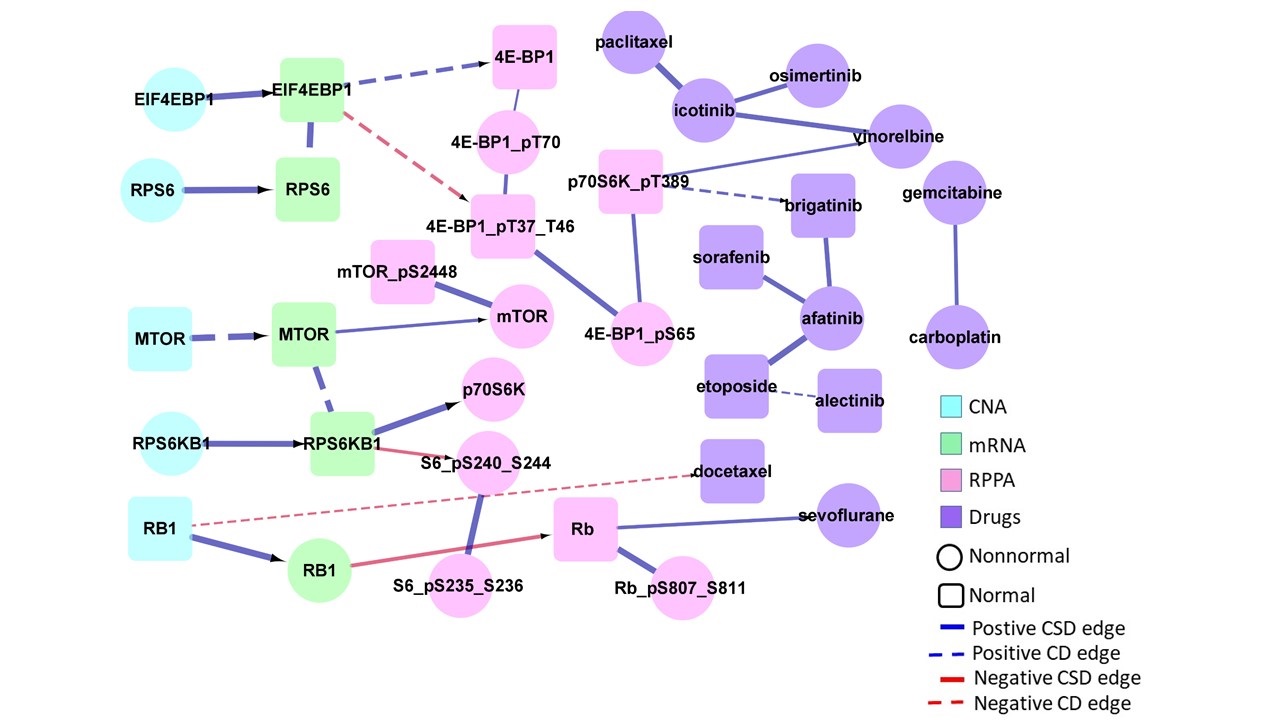}
     \caption{The estimated multilayered network for the TSC/mTOR pathway. Blue and red edges indicate positive and negative dependencies, while CD and CSD stand for conditionally dependent and conditionally sign-dependent edges respectively. The width of the edges is proportional to the posterior inclusion probabilities.}
     \label{fig:graph_tscmtor}
 \end{figure}

\begin{figure}
     \centering
     \includegraphics[scale=0.55]{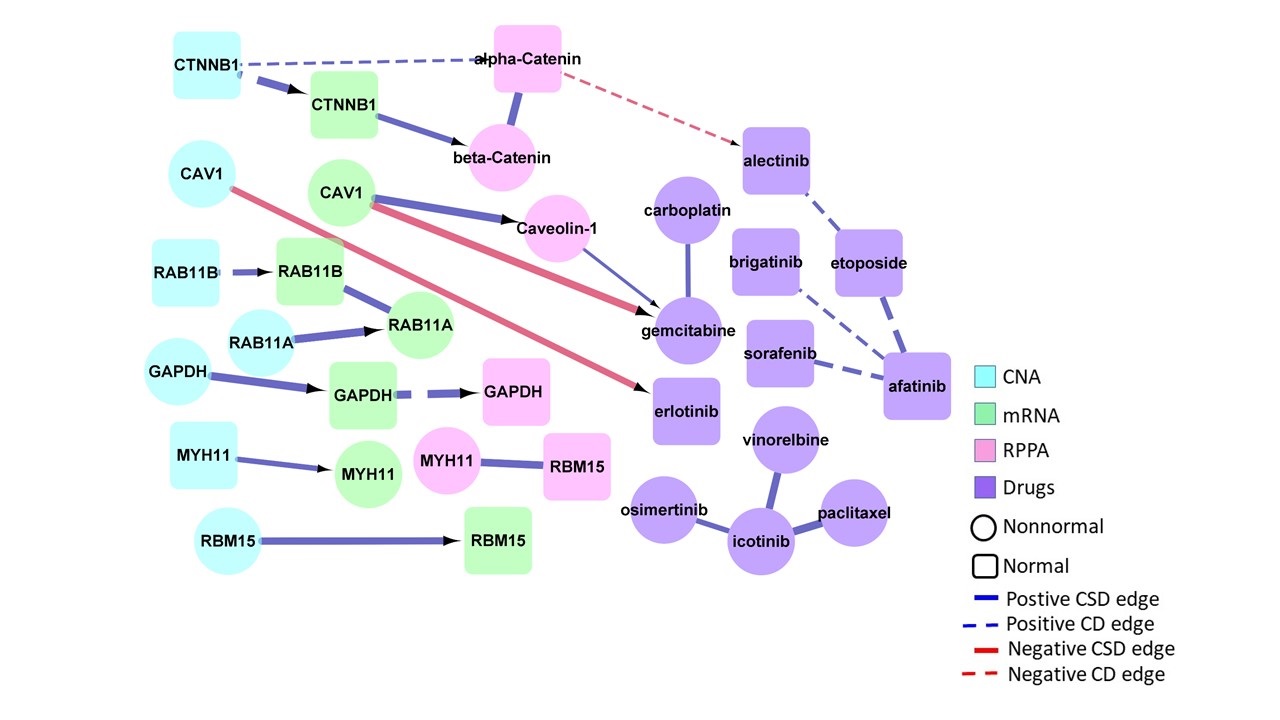}
     \caption{The estimated multilayered network for the breast reactive pathway. Blue and red edges indicate positive and negative dependencies, while CD and CSD stand for conditionally dependent and conditionally sign-dependent edges respectively. The width of the edges is proportional to the posterior inclusion probabilities.}
     \label{fig:graph_breastreactive}
 \end{figure}

\begin{figure}
     \centering
     \includegraphics[scale=0.55]{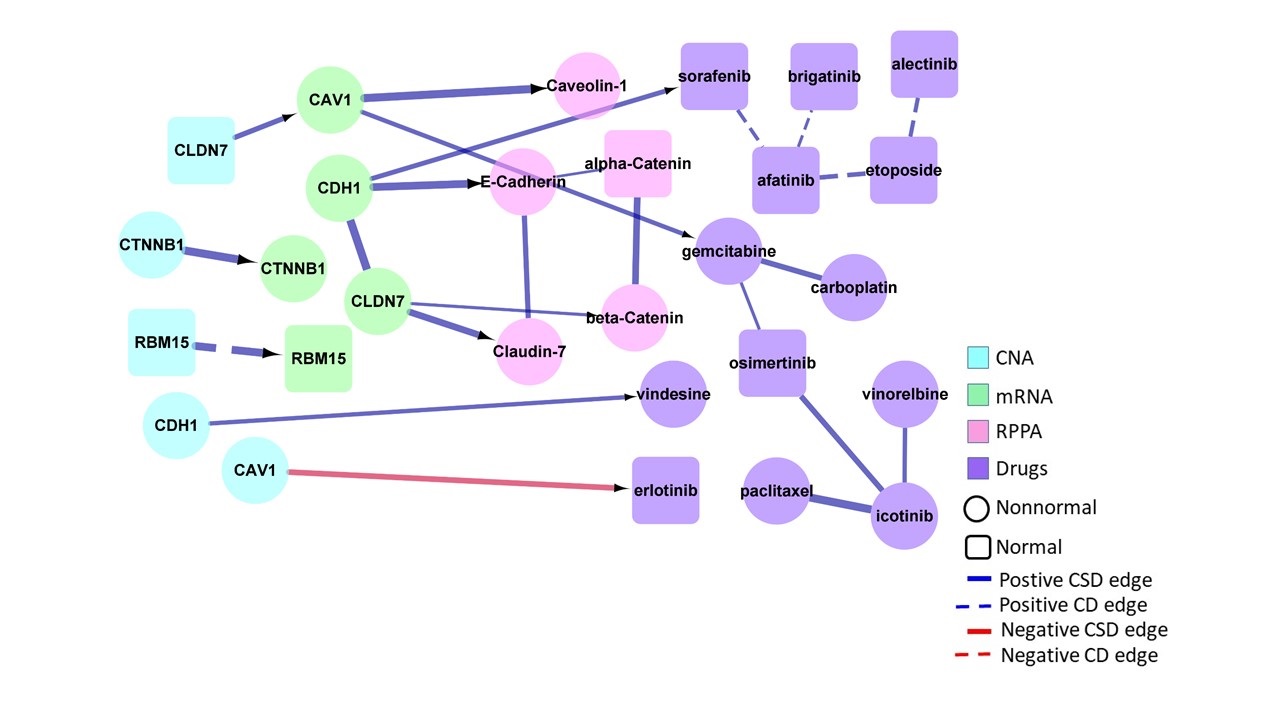}
     \caption{The estimated multilayered network for the core reactive pathway. Blue and red edges indicate positive and negative dependencies, while CD and CSD stand for conditionally dependent and conditionally sign-dependent edges respectively. The width of the edges is proportional to the posterior inclusion probabilities.}
     \label{fig:graph_corereactive}
 \end{figure}

 We have pathway-level graphs obtained using RCGM for $10$ key signaling pathways: apoptosis, cell cycle, DNA damage response, EMT, RAS/MAPK, RTK, PI3K/AKT, TSC/mTOR, core reactive and breast reactive. The graph for DNA damage response is displayed in Figure~\ref{fig: DDR graph}. The rest of the pathway graphs are displayed in Figures ~\ref{fig:graph_apoptosis} -- \ref{fig:graph_corereactive}. For each graph, the edge width is proportional to the posterior edge inclusion probability obtained from MCMC samples. The platform-wise violin plot of the estimated non-normality scores $\pi_v$ are displayed in Figure~\ref{fig:Estimated pi violin plot}.
 
\begin{figure}[t!]
    \centering
    \includegraphics[width=\textwidth]{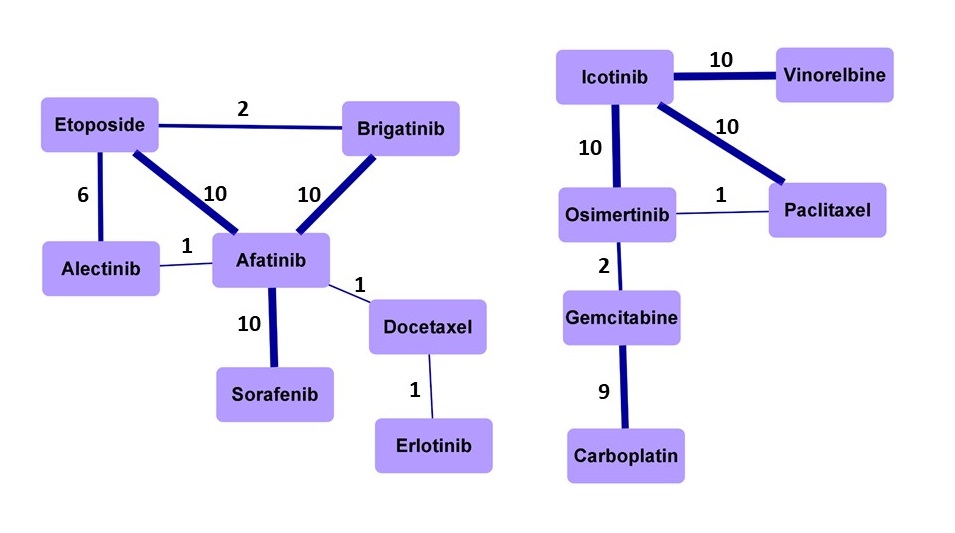}
    \caption{Dependent drug pairs captured across the $10$ pathways. Each edge width is proportional to, and marked with the number of pathways (out of $10$) the corresponding drug-drug dependency is detected in.}
    \label{fig: drug-drug interactions 0}
\end{figure}

We display the between-drug edges detected across pathways in Figure~\ref{fig: drug-drug interactions 0}. Two clusters of drugs emerge from this analysis - the first cluster consisting of icotinib, vinorelbine, paclitexel, osimertinib,  gemcitabine and carboplatin, and the second cluster including afatinib, brigatinib, etoposide, sorafenib, alectinib, docetaxel and erlotinib. Dependencies of icotinib with vinorelbine, paclitaxel and osimertinib and those of afatinib with sorafenib, etoposide and brigatinib were detected in all $10$ pathways. Strong dependencies between drugs indicate that drugs' individual effects on cell viability after scaling for robustness and adjusting for effects of all other variables are strongly positively correlated.
 \begin{figure}
     \centering
     \includegraphics[width=0.49\textwidth]{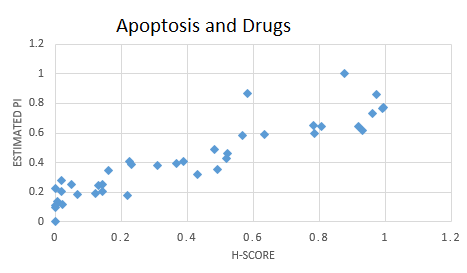}
     \includegraphics[width=0.49\textwidth]{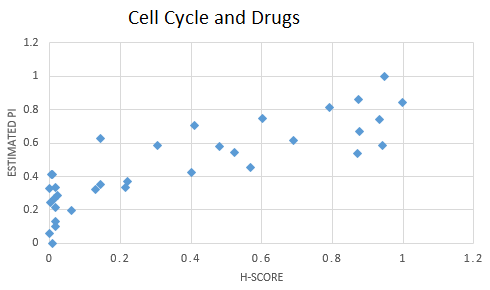}
      \includegraphics[width=0.49\textwidth]{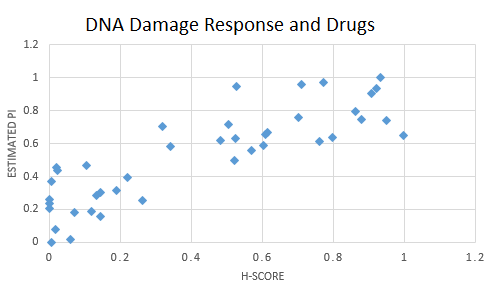}
       \includegraphics[width=0.49\textwidth]{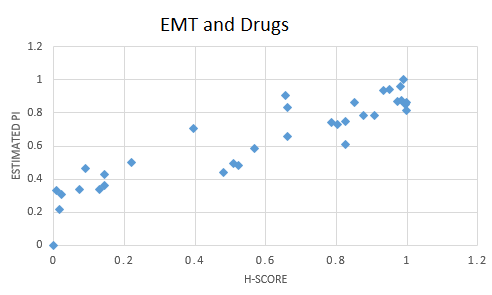}
        \includegraphics[width=0.49\textwidth]{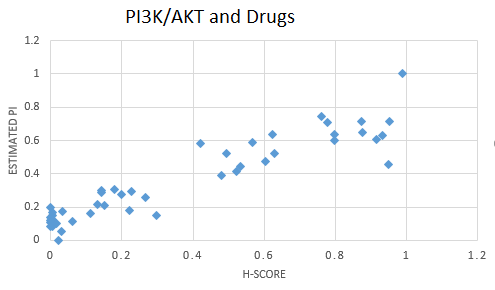}
         \includegraphics[width=0.49\textwidth]{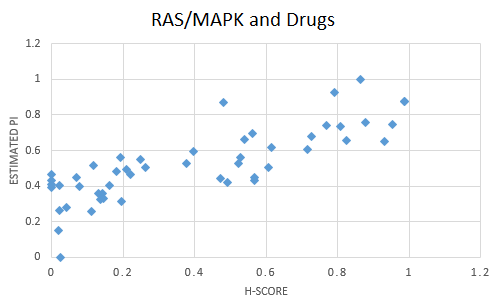}
     \includegraphics[width=0.49\textwidth]{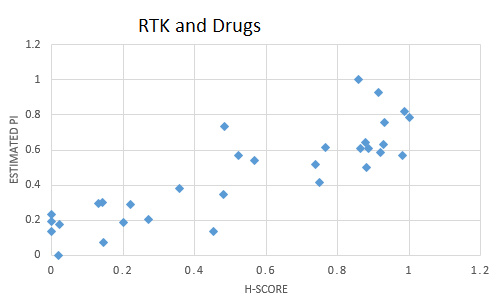}
      \includegraphics[width=0.49\textwidth]{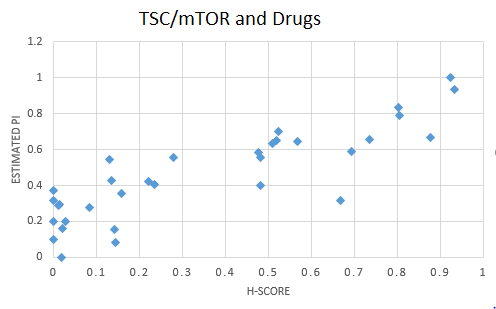}
       \includegraphics[width=0.49\textwidth]{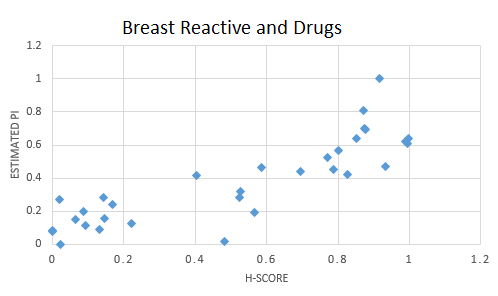}
      \includegraphics[width=0.49\textwidth]{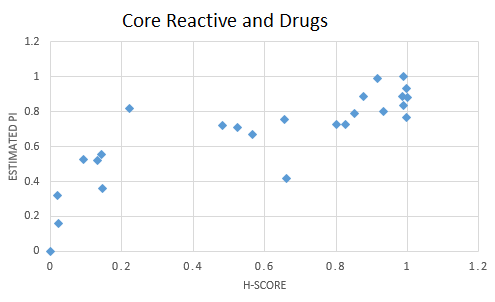}
     \caption{Pathway-wise H-score vs. $\hat{\pi}$}
     \label{fig:nnscore_cellcycle}
 \end{figure}
\clearpage
\section{Supplementary Tables}
We display the names of pathways chosen in our study along with the count of gene and protein memberships in each in Table~\ref{tab: count of nodes across pathways}. The list of genes and proteins corresponding to each pathway is given in Table~\ref{tab:antibody-gene pathway wise list}. Tables~\ref{tab: drug mech1} -- \ref{tab: drug mech2} display the target genes, mechanism of action and primary diseases associated with the drugs used in our study.
\begin{table}[ht]
\centering
\begin{tabular}{llll}
  \hline
Pathway & CNA & mRNA & RPPA \\ 
  \hline
Apoptosis & 9 & 9 & 10 \\ 
  Cell cycle & 7 & 7 & 9 \\ 
  DNA damage response & 10 & 10 & 13 \\ 
  EMT & 7 & 7 & 9 \\ 
  PI3K/AKT & 10 & 10 & 15 \\ 
  RAS/MAPK & 10 & 10 & 19 \\ 
  RTK & 5 & 5 & 11 \\ 
  TSC/mTOR & 5 & 5 & 12 \\ 
  Breast reactive & 7 & 7 & 7 \\ 
  Core reactive & 5 & 5 & 7 \\ 
   \hline
\end{tabular}
\caption{Number of nodes in each platform of genomic information across 10 pathways. }%
\label{tab: count of nodes across pathways}
\end{table}
\begin{table}[ht]
\centering
\begin{tabular}{lll}
  \hline
   \hline
Pathway & Gene & Antibodies \\
  \hline
Apoptosis & BAD & Bad\_pS112 \\ 
  Apoptosis & BAK1 & Bak\_Caution \\ 
  Apoptosis & BAX & Bax \\ 
  Apoptosis & BCL2 & Bcl-2 \\ 
  Apoptosis & BCL2L1 & Bcl-xL \\ 
  Apoptosis & BID & Bid\_Caution \\ 
  Apoptosis & BCL2L11 & Bim(CST2933), Bim(EP1036) \\ 
  Apoptosis & CASP7 & Caspase-7\_cleavedD198\_Caution \\ 
  Apoptosis & BIRC2 & cIAP\_Caution \\ 
   \hline
   Cell cycle & CDK1 & CDK1 \\ 
  Cell cycle & CCNB1 & Cyclin\_B1 \\ 
  Cell cycle & CCNE1 & Cyclin\_E1 \\ 
  Cell cycle & CCNE2 & Cyclin\_E2\_Caution \\
   Cell cycle & FOXM1 & FoxM1 \\ 
  Cell cycle & CDKN1B & p27, p27\_pT157\_Caution, p27\_pT198 \\ 
  Cell cycle & PCNA & PCNA\_Caution \\
  \hline
  DNA damage response & TP53BP1 & 53BP1 \\ 
  DNA damage response & ATM & ATM \\ 
  DNA damage response & BRCA2 & BRCA2\_Caution \\ 
  DNA damage response & CHEK1 & Chk1\_Caution, Chk1\_pS345\_Caution \\ 
  DNA damage response & CHEK2 & Chk2, Chk2\_pT68\_Caution \\ 
  DNA damage response & XRCC5 & Ku80\_Caution \\ 
  DNA damage response & MRE11A & Mre11\_Caution \\ 
  DNA damage response & TP53 & p53\_Caution \\
  DNA damage response & RAD50 & RAD50 \\ 
  DNA damage response & RAD51 & RAD51 \\ 
  DNA damage response & XRCC1 & XRCC1\_Caution \\
    \hline
EMT & CTNNB1 & alpha-Catenin, beta-Catenin,\\
&& beta-Catenin\_pT41\_S45 \\ 
  EMT & CLDN7 & Claudin-7 \\ 
  EMT & COL6A1 & Collagen\_VI \\ 
  EMT & CDH1 & E-Cadherin \\ 
  EMT & FN1 & Fibronectin \\ 
  EMT & CDH2 & N-Cadherin \\ 
  EMT & SERPINE1 & PAI-1 \\ 
     \hline
   \end{tabular}
\end{table}
    
    \begin{table}[ht]
\centering
\begin{tabular}{lll}
  \hline
  PI3K/AKT & AKT1 & Akt, Akt\_pS473, Akt\_pT308 \\ 
  PI3K/AKT & AKT2 & Akt, Akt\_pS473, Akt\_pT308 \\ 
  PI3K/AKT & AKT3 & Akt, Akt\_pS473, Akt\_pT308\\
  PI3K/AKT & GSK3B &  GSK-3-beta\_Caution, GSK3-alpha-beta,\\
   & & 
  GSK3-alpha-beta\_pS21\_S9, GSK3\_pS9\\
  PI3K/AKT & GSK3A & GSK3\_pS9, GSK3-alpha-beta,\\
   & & GSK3-alpha-beta\_pS21\_S9\\ 
  PI3K/AKT & INPP4B & INPP4B \\ 
  PI3K/AKT & CDKN1B & p27, p27\_pT157\_Caution, p27\_pT198 \\ 
  PI3K/AKT & AKT1S1 & PRAS40\_pT246 \\ 
  PI3K/AKT & PTEN & PTEN \\ 
  PI3K/AKT & TSC2 & Tuberin, Tuberin\_pT1462 \\ 
    \hline
RAS/MAPK & ARAF & A-Raf\_pS299\_Caution \\ 
  RAS/MAPK & JUN & c-Jun\_pS73 \\ 
  RAS/MAPK & RAF1 & C-Raf(BD610151)\_Caution, C-Raf(MP05-739),\\
  && C-Raf\_pS338 \\ 
  RAS/MAPK & MAPK1 & ERK2\_Caution, MAPK\_pT202\_Y204 \\ 
  RAS/MAPK & MAPK8 & JNK\_pT183\_Y185 \\ 
  RAS/MAPK & MAPK3 & MEK1 \\ 
  RAS/MAPK & MAP2K1 & MEK1\_pS217\_S221, p38 alpha MAPK \\ 
  RAS/MAPK & MAPK14 & p38\_MAPK, p38\_pT180\_Y182, p90RSK\_Caution \\ 
  RAS/MAPK & RPS6KA1 & p90RSK\_pT359\_S363\_Caution, p90RSK\_pT573\_Caution,\\
  && RSK1-2-3\_Caution, YB-1 \\ 
  RAS/MAPK & YBX1 & c-Jun\_pS73, C-Raf(BD610151)\_Caution \\ 
   \hline
   RTK & EGFR & EGFR, EGFR\_pY1068\_Caution, EGFR\_pY1173 \\ 
  RTK & ERBB2 & HER2, HER2\_pY1248\_Caution \\ 
  RTK & ERBB3 & HER3, HER3\_pY1289\_Caution \\ 
  RTK & SHC1 & Shc\_pY317 \\ 
  RTK & SRC & Src, Src\_pY416\_Caution, Src\_pY527 \\ 
   \hline
   TSC/mTOR & EIF4EBP1 & 4E-BP1, 4E-BP1\_pS65, 4E-BP1\_pT37\_T46,\\
   && 4E-BP1\_pT70 \\ 
  TSC/mTOR & MTOR & mTOR, mTOR\_pS2448\_Caution \\ 
  TSC/mTOR & RPS6KB1 & p70S6K, p70S6K\_pT389 \\ 
    TSC/mTOR & RB1 & Rb\_Caution, Rb\_pS807\_S811 \\ 
  TSC/mTOR & RPS6 & S6\_pS235\_S236, S6\_pS240\_S244 \\ 
   \hline
      \end{tabular}
\end{table}
    
    \begin{table}[ht]
\centering
\begin{tabular}{lll}
\hline
   Breast reactive & CTNNB1 & alpha-Catenin, beta-Catenin, beta-Catenin\_pT41\_S45 \\ 
  Breast reactive & CAV1 & Caveolin-1 \\ 
  Breast reactive & GAPDH & GAPDH\_Caution \\ 
  Breast reactive & MYH11 & MYH11 \\ 
  Breast reactive & RBM15 & RBM15 \\ 
   \hline
   Core reactive & CTNNB1 & alpha-Catenin, beta-Catenin, beta-Catenin\_pT41\_S45 \\ 
  Core reactive & CAV1 & Caveolin-1 \\ 
  Core reactive & CLDN7 & Claudin-7 \\ 
  Core reactive & CDH1 & E-Cadherin \\ 
  Core reactive & RBM15 & RBM15 \\ 
   \hline
   \end{tabular}
   \caption{Antibodies and their target genes considered in RPPA analysis}
   \label{tab:antibody-gene pathway wise list}
\end{table}
\begin{table}[ht]
\centering
\begin{tabular}{c|p{4cm}|p{4cm}|p{3cm}}
  \hline
 Drug & Mechanism of Action & Target genes & Indication \\
  \hline
  Icotinib & EGFR inhibitor & EGFR &  NSCLC \\ 
  Osimertinib & EGFR inhibitor & EGFR &  NSCLC \\ 
  Gefitinib & EGFR inhibitor & EGFR &  NSCLC \\ 
  Afatinib & EGFR inhibitor & EGFR,ERBB2,ERBB4 &  NSCLC \\
   Erlotinib & EGFR inhibitor & EGFR, NR1I2 &  NSCLC, pancreatic cancer \\ 
   Brigatinib & EGFR inhibitor, ALK tyrosine kinase receptor inhibitor & ALK, EGFR &  NSCLC \\ Alectinib & ALK tyrosine kinase receptor inhibitor & ALK, MET &  NSCLC \\
  Ceritinib & ALK tyrosine kinase receptor inhibitor & ALK, FLT3, IGF1R, INSR, TSSK1B &  NSCLC \\ 
  Crizotinib & ALK tyrosine kinase receptor inhibitor & ALK, MET &  NSCLC \\ 
  Paclitaxel & tubulin polymerization inhibitor & BCL2,MAP2,MAP4, MAPT,NR1I2,TLR4, TUBB,TUBB1 & ovarian and breast cancer,  NSCLC \\ 
  Docetaxel & tubulin polymerization inhibitor & BCL2, MAP2, MAP4, MAPT,NR1I2, TUBB, TUBB1 & NSCLC, breast and prostate cancer, gastric adenocarcinoma, HNSCC \\
  Vindesine & tubulin polymerization inhibitor & TUBB, TUBB1 & NSCLC,melanoma, breast cancer \\
   Vinorelbine & tubulin polymerization inhibitor & TUBA1A, TUBA1B, TUBA1C, TUBA3C, TUBA3D, TUBA3E, TUBA4A, TUBB, TUBB1, TUBB2A, TUBB2B, TUBB3, TUBB4A, TUBB4B, TUBB6, TUBB8 &  NSCLC \\ 
  Etoposide & topoisomerase inhibitor & TOP2A, TOP2B &  NSCLC \\ 
 \hline\\
  \end{tabular}
  \caption{List of drugs used in our analysis along with their targets.}
  \label{tab: drug mech1}
  \end{table}
  \begin{table}[ht]
\centering
\begin{tabular}{c|p{4cm}|p{4cm}|p{3cm}}
  \hline
 Drug & Mechanism of action & Target genes & Indication \\
  \hline
  Pemetrexed & dihydrofolate reductase inhibitor, thymidylate synthase inhibitor & ATIC, DHFR, GART, TYMS &  NSCLC, mesothelioma \\ 
  Gemcitabine & ribonucleotide reductase inhibitor & CMPK1,RRM1,RRM2, TYMS &  NSCLC,ovarian, breast and pancreatic cancer \\
  Sevoflurane & membrane integrity inhibitor & ATP2C1, ATP5D, GABRA1, GABRA2, GABRA3, GABRA4, GABRA5, GABRA6, GABRB1, GABRB2, GABRB3, GABRD, GABRE, GABRG1, GABRG2, GABRG3, GABRP, GABRQ, GLRA1,GLRB,GRIA1, KCNA1, KCNK10, KCNK18,KCNK2, KCNK3, KCNK9,MT-ND1 & anesthetic \\ 
    Cisplatin & DNA synthesis inhibitor, DNA alkylating agent & XIAP & testicular carcinoma, ovarian and bladder cancer \\ 
  Carboplatin & DNA alkylating agent, DNA inhibitor &  & ovarian cancer \\ 
  Sorafenib & FLT3 inhibitor, KIT inhibitor, PDGFR tyrosine kinase receptor inhibitor, RAF inhibitor, RET tyrosine kinase inhibitor, VEGFR inhibitor & BRAF,DDR2,FGFR1, FLT3,FLT4,KDR,KIT, PDGFRB, RAF1, RET, FLT1 & renal cell carcinoma (RCC), thyroid cancer, hepatocellular carcinoma (HCC) \\ 
  \hline
  \end{tabular}
  \caption{List of drugs used in our analysis along with their targets.}
  \label{tab: drug mech2}
  \end{table}


\end{document}